\documentclass[a4paper,fleqn]{cas-sc}

\usepackage[authoryear]{natbib}

\usepackage{subcaption}
\usepackage{xcolor}
\usepackage{soul}
\PassOptionsToPackage{colorlinks=true}{hyperref}
\usepackage{cleveref}
\usepackage{makecell}
\usepackage{fancyvrb}
\usepackage{graphicx}
\usepackage{longtable}
\usepackage{enumitem}
\usepackage{xparse}    % 仅用于定义带默认值的命令（可选）
\usepackage{amsmath,amssymb}
\usepackage{booktabs}



\hypersetup{
    colorlinks = true,    % 启用彩色链接
    linkcolor = blue,     % 内部链接（如目录、章节引用）
    citecolor = blue,     % 文献引用
    urlcolor = blue,      % URL链接
    filecolor = blue,     % 文件链接
    allcolors = blue,     % 强制所有链接为蓝色（可选）
}

\crefname{figure}{\textcolor{blue}{Fig.}}{\textcolor{blue}{Figs.}} % 单数和复数形式
\crefname{table}{\textcolor{blue}{Table}}{\textcolor{blue}{Tables}}
\crefname{equation}{\textcolor{blue}{Eq.}}{\textcolor{blue}{Eqs.}}
\crefname{algorithm}{\textcolor{blue}{algorithm}}{\textcolor{blue}{algorithms}}
\Crefname{algorithm}{\textcolor{blue}{Algorithm}}{\textcolor{blue}{Algorithms}}

\def\tsc#1{\csdef{#1}{\textsc{\lowercase{#1}}\xspace}}
\tsc{WGM}
\tsc{QE}
\tsc{EP}
\tsc{PMS}
\tsc{BEC}
\tsc{DE}
\begin{document}
\let\WriteBookmarks\relax
\def\floatpagepagefraction{1}
\def\textpagefraction{.001}

\shorttitle{Mobility Behavior and Epidemic Dynamics under Policy Change}

\shortauthors{Ye et~al.}

\title [mode = title]{From Urban Mobility to Epidemic Dynamics: A Mixture-of-Experts Framework with Preference Alignment for Policy Scenario Simulation}                      
% \tnotemark[1,2]

% \tnotetext[1]{This document is the results of the research
%    project funded by the National Science Foundation.}

% \tnotetext[2]{The second title footnote which is a longer text matter
%    to fill through the whole text width and overflow into
%    another line in the footnotes area of the first page.}

% Hide author information for Double anonymized peer review-----------------
\author[1,2]{Yun Ye}

% \fnmark[†]

%\ead{yeyun@connect.hku.hk}

% \credit{Conceptualization of this study, Methodology, Software}

\affiliation[1]{organization={Centre for Global Infrastructure Resilience, The Bartlett School of Sustainable Construction, University College London},
    %addressline={}, 
    city={London},
    % citysep={}, % Uncomment if no comma needed between city and postcode
    postcode={WC1E 7HB}, 
    country={United Kingdom}}

\affiliation[2]{organization={SpaceTimeLab, Department of Civil, Environmental and Geomatic Engineering, University College London},
    %addressline={}, 
    city={London},
    % citysep={}, % Uncomment if no comma needed between city and postcode
    postcode={WC1E 6BT}, 
    country={United Kingdom}}
    
% Second 
\author[3,2]{Arsalan Dezhkam}
% Footnote of the second author
% \fnmark[†]

\affiliation[3]{organization={Queen Square Institute of Neurology, University College London},
    %addressline={}, 
    city={London},
    % citysep={}, % Uncomment if no comma needed between city and postcode
    postcode={WC1N 3BG}, 
    country={United Kingdom}
    }
    
% Third
\author[2]{Junyuan Liu}
    
% Fourth
\author[2]{Xinglei Wang}

% Fifth author
\author[2]{Tao Cheng}[orcid=0000-0002-5503-9813]
% Footnote of the second author
% Email id
\ead{tao.cheng@ucl.ac.uk}
\cormark[1]

\cortext[cor1]{Corresponding author}

% \footnote{These authors contributed equally to this work.}
% \fntext[†]{These authors contributed equally to this work.}
%-------------------------------------------------------------------

% \nonumnote{This note has no numbers. In this work we demonstrate $a_b$
%   the formation Y\_1 of a new type of polariton on the interface
%   between a cuprous oxide slab and a polystyrene micro-sphere placed

\begin{abstract}
Non-pharmaceutical interventions (NPIs) alter epidemic risk through behavioral reallocations, not simply aggregate mobility reductions. Scenario-based NPI analysis therefore requires a behavioral layer that translates alternative policy calendars into plausible activity and mobility trajectories before downstream outcomes are simulated. We introduce UrbanShare-MoE-PA, a data-driven agent-level framework that maps factual and alternative NPI calendars to daily time-allocation trajectories and propagates them through a calibrated behavior-driven SEIR simulator. The behavioral engine decomposes each agent-day into travel share, POI-category allocation conditional on staying, and travel-mode allocation conditional on traveling. It combines a structured UrbanShare baseline, mixture-of-experts heads for heterogeneous POI and mode responses, and phase-aware preference alignment for calendar-conditioned rollouts. Using data from 911 agents in Singapore observed from March to August 2020, we evaluate factual reconstruction, four alternative lockdown calendars, and epidemic-activity trade-offs. UrbanShare-MoE improves POI reconstruction over the baseline, while UrbanShare-MoE-PA achieves the lowest travel-mode errors and the clearest alternative-calendar trajectories. In the calibrated SEIR simulation, early lockdown lowers infectious burden, late lockdown increases it, and short lockdown preserves the highest weighted activity with only a modest increase in epidemic burden relative to the original policy. These results show that epidemic-activity conclusions depend on how policy calendars are translated into behavior, and that agent-level mobility-share modeling provides an interpretable bridge between policy timing, behavior, and downstream simulation.
\end{abstract}

% \includegraphics{figs/grabs.pdf}

% \begin{highlights}
% \item Agent-level time allocation links policy calendars to mobility trajectories.
% \item Phase-conditioned MoE captures policy shifts and heterogeneous behavioral regimes.
% \item Preference alignment improves rollout coherence near policy transitions.
% \item A calibrated epidemic simulator compares alternative lockdown calendars.
% \item Sector weights reveal epidemic-activity trade-offs across scenarios.
% \end{highlights}

% Keywords
% Each keyword is separated by \sep

\maketitle

% introduction.tex

\section{Introduction}
\label{sec:introduction}

The COVID-19 pandemic exposed a fundamental tension in urban and transport
policy. Dense cities rely on frequent human movement, public transport,
activity centers, and repeated daily routines, yet the same connectivity can
facilitate infectious-disease transmission. Governments therefore adopted
non-pharmaceutical interventions (NPIs), including social distancing, working
from home, travel restrictions, and lockdowns, to reduce contact opportunities.
Such interventions, however, affect more than aggregate mobility: they alter
where people spend time, how they travel, and which urban activities remain accessible. For policy analysis, the relevant question is therefore not simply whether restrictions reduce movement, but how changes in the timing and duration of interventions reshape daily behavior, exposure, and retained urban activity.

Transportation research provides an important behavioral foundation for this
problem. Daily mobility is not a random sequence of trips, but a structured
allocation of limited time across activity participation, destinations, travel
modes, and repeated routines \citep{timmermans2002time, cagney2020urban}.
Empirical studies further show that individuals tend to move within relatively
regular and bounded activity spaces
\citep{Gonzlezunderstanding2008, Songmodelling2010, Sununderstanding2013,
Hasanunderstanding2013}. During COVID-19, these structures were reorganized
rather than uniformly suppressed. Travel declined, home-based time increased,
public-transport use changed, and visits to commercial, recreational, and
service destinations shifted across intervention phases
\citep{Beckinsights2020, peng2025understanding, yabe2020non,
lucchini2021living}. Longer-term studies similarly report persistent changes in
telecommuting, shopping, travel timing, and home-centered routines
\citep{restrepo2022work, javadinasr2021enduring, salon2022covid,
aaditya2023long, huang2023travel, shi2024long, shi2024year,
shi2026dissimilarities}. These findings suggest that responses to NPIs are more
appropriately represented as a \emph{daily time-allocation process} than as a
single reduction in mobility volume.

This behavioral structure is also important for epidemic modeling because
infection opportunities depend on where and how time is spent. Classical
compartmental models provide the foundation for epidemic dynamics
\citep{Kermacka1927, Hethcotethe2000}, while network and metapopulation models
represent contact structure, spatial interaction, and travel-related exposure
\citep{Keelingnetworks2005, Salatha2010, Wanginferring2018,
Kniplepidemic2013}. Transport-oriented models have further incorporated public
transport, commuting, urban activity zones, and agent-level movement into
disease-spread simulation
\citep{Xuspatial2013, Meisimulating2015, Perezan2009, Zhangincluding2015,
liu2022modelling}. In Singapore, for example, \citet{liu2022modelling}
incorporated area-based and MRT-related exposure into an SEIR framework for
evaluating pandemic control measures. Collectively, these studies establish
that mobility and activity locations are not peripheral inputs to epidemic
dynamics; they form part of the exposure mechanism itself. Meaningful policy
scenario analysis therefore requires not only a downstream epidemic model, but
also a credible behavioral representation upstream of it.

A key limitation arises when this modeling chain is used for alternative
policy scenarios. Most epidemic-policy models treat mobility as an observed
input, a flow matrix, or a contact multiplier. Such representations are
appropriate for retrospective analysis, where the realized mobility pattern is
already known. Under an alternative policy calendar, however, the behavioral
input itself becomes unobserved. If a lockdown had started one week earlier,
ended two weeks sooner, or remained in place for longer, the corresponding
daily allocation of time across home, other activity locations, and travel modes
would not be directly available. Aggregate mobility indicators cannot resolve
this problem because two days with similar total movement may involve very
different allocations across retail, workplaces, public transport, private
vehicles, and home, with correspondingly different implications for exposure
and retained activity. Although mobile-phone mobility studies have established
strong associations between mobility reduction and transmission change
\citep{buckee2020aggregated, kraemer2020effect, bryant2020estimating}, the
interpretive limitations of aggregate mobility proxies remain important
\citep{kishore2022evaluating, wardle2023gaps}. The upstream problem is therefore to generate behaviorally plausible daily trajectories under a modified policy calendar before their epidemic or activity consequences are compared.

Constructing such a behavioral generator introduces several modeling
challenges. First, a policy regime should not be treated merely as another
covariate. Similar individual and contextual conditions may lead to different
behavioral responses under different levels of restriction, requiring the
policy phase to explicitly modulate the behavioral representation. Second,
responses remain heterogeneous even within the same policy regime. Individuals
differ in their activity needs, habitual mobility, transport dependence, and
capacity to substitute out-of-home activities with home-based alternatives;
a single prediction pathway may therefore average over distinct behavioral
regimes. Third, accurate reconstruction under the factual calendar does not
necessarily guarantee coherent simulation when known policy regimes are moved
to different calendar positions. Alternative-calendar simulation requires
recursive rollout, in which previously generated behavior becomes part of the
subsequent behavioral history. Errors around policy transitions can therefore
persist or accumulate, making phase-consistent adjustment particularly
important when intervention boundaries are shifted.

These challenges motivate a behavioral-policy framework with three
complementary capabilities: structured daily allocation, heterogeneous
phase-conditioned response modeling, and stable rollout under modified policy
calendars. The framework focuses on \emph{calendar-conditioned policy scenario
simulation}, in which policy regimes represented in the observed data are
shifted or extended in time and the corresponding behavioral trajectories are
regenerated. Its generalization target is the temporal recombination and
duration adjustment of observed policy regimes.

To address this problem, this study proposes
\textbf{UrbanShare-MoE-PA}, a data-driven agent-level framework that translates
factual and alternative policy calendars into structured daily behavioral
trajectories. The behavioral representation factorizes each agent-day into
three linked components: the fraction of time spent traveling, the allocation
of non-travel time across POI categories, and the allocation of travel time
across transport modes. Building on this structure, Feature-wise Linear
Modulation (FiLM) explicitly conditions the latent context on the current
policy phase, while sparse Top-$K$ Mixture-of-Experts (MoE) heads capture
heterogeneous POI and mode-response regimes. A phase-aware preference alignment
stage then fine-tunes the behavioral policy by ranking phase-consistent
actions above random and wrong-phase alternatives, with greater emphasis near
policy-transition boundaries. The alignment targets the temporal and phase consistency of recursive
behavioral rollout when the policy calendar is shifted.

The generated trajectories are subsequently propagated through a
behavior-driven SEIR simulator in which exposure is linked to POI-category and
travel-mode time, and through a sector-weighted activity index constructed from
the same behavioral representation. Previous COVID-19 policy studies have
shown that intervention timing, compliance, and duration can substantially
affect epidemic outcomes
\citep{Premthe2020, paperquantifying2020, Davieseffects2020,
Hossainthe2020, Kretzschmarimpact2020, Dingtlqp2021,
Sharmaunderstanding2021}, while restrictions also impose uneven social and
economic costs across urban functions and sectors
\citep{bonaccorsi2020economic, Beckinsights2020}. By generating the behavioral
trajectory before evaluating these downstream outcomes, the proposed framework
provides a common behavioral basis for comparing alternative policy calendars
rather than imposing mobility changes directly on the epidemic model.

The main contributions of this paper are summarized as follows:
\begin{itemize}

    \item We formulate alternative-calendar NPI analysis as a
    \emph{data-driven agent-level policy scenario simulation} problem. Instead
    of treating mobility as an exogenous aggregate input, the framework first
    generates the daily behavioral trajectory associated with each policy
    calendar and then propagates it to downstream outcomes.

    \item We develop a structured daily time-allocation model that jointly
    predicts travel fraction, POI-category allocation, and travel-mode
    allocation. FiLM-based phase modulation explicitly represents
    policy-conditioned behavioral shifts, while sparse Top-$K$ MoE heads
    capture heterogeneous allocation regimes across agent-days.

    \item We introduce phase-aware preference alignment for
    alternative-calendar rollout. By ranking observed and phase-consistent
    actions above perturbed and wrong-phase alternatives, with greater emphasis
    near policy boundaries, the aligned model is designed to preserve
    behavioral coherence when observed policy regimes are shifted in time.

    \item We link the generated behavioral trajectories to a behavior-driven
    SEIR simulator and a sector-weighted urban activity index, enabling
    internally consistent comparison of epidemic burden and retained activity
    under early, late, short, and long lockdown calendars.

\end{itemize}

The remainder of the paper is organized as follows.
Section~\ref{sec:dataset} describes the mobility dataset and the construction
of structured daily behavioral profiles. Section~\ref{sec:methodology}
introduces UrbanShare-MoE-PA, phase-aware preference alignment,
calendar-conditioned simulation, epidemic dynamics, and the scenario evaluation
framework. Section~\ref{sec:results} presents the behavioral and downstream
simulation results. Section~\ref{sec:discussion} discusses the findings,
implications, limitations, and future research directions.
Section~\ref{sec:conclusion} concludes the paper.

\section{Dataset}
\label{sec:dataset}

\begin{figure}[h]
\centering
\includegraphics[scale=.25]{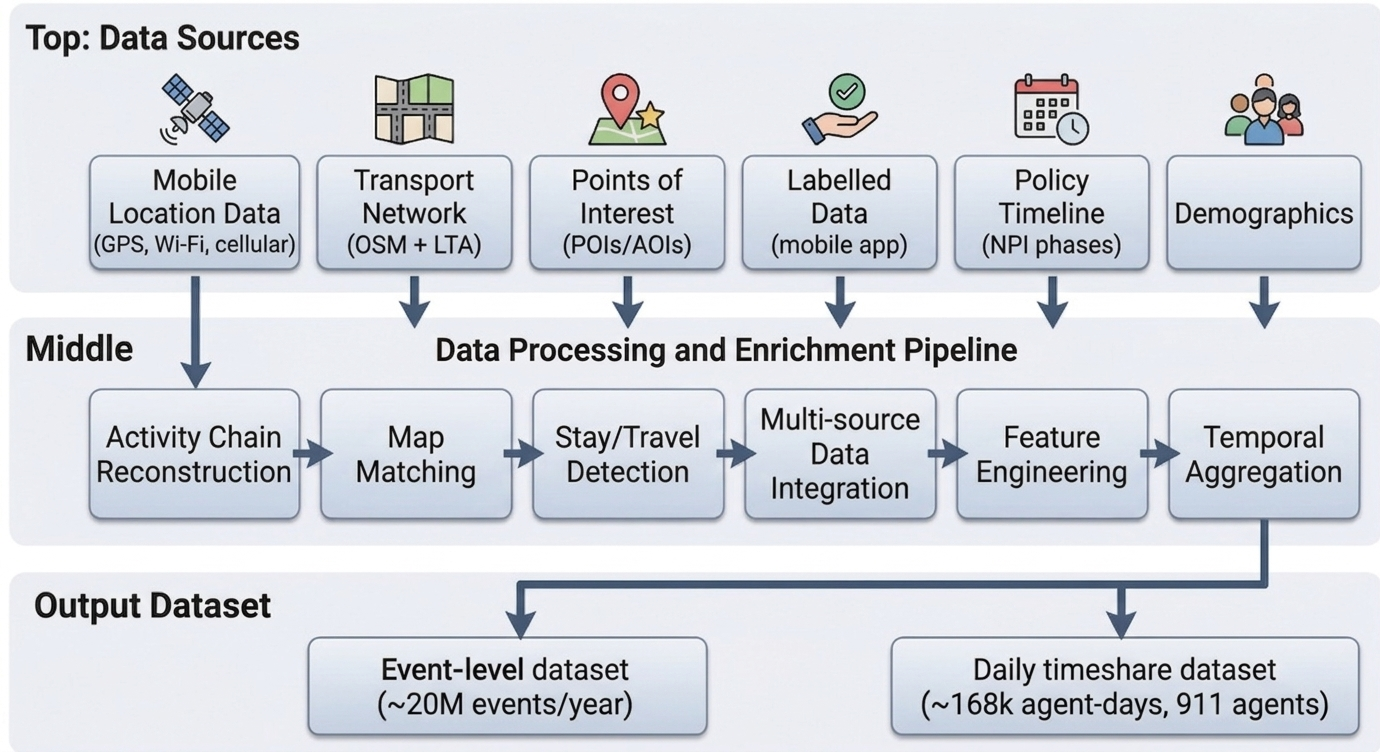}
\caption{Data construction pipeline.}
\label{FIG:data framework}
\end{figure}

To support behaviorally grounded policy scenario analysis, we construct a
unified data framework that transforms real-world individual travel trajectory data
into structured daily activity and mobility representations. As illustrated in
Figure~\ref{FIG:data framework}, the framework integrates individual-level
spatio-temporal trajectories with transport networks, points of interest (POIs),
demographic attributes, and non-pharmaceutical intervention (NPI) policy timelines.
These data sources are processed through a multi-stage enrichment pipeline to
generate both event-level activity chains and daily behavioral profiles for model
training, preference alignment, calendar-conditioned policy scenario simulation, and downstream
epidemic evaluation.

At a high level, raw spatio-temporal trajectories are first converted into activity
chains through map matching and stay--travel segmentation. These intermediate
representations are then enriched using semantic, spatial, demographic, and
policy-related information from external data sources. Finally, temporal aggregation
and feature engineering are performed to construct daily behavior profiles that
jointly describe activity allocation, travel mode allocation, and mobility dynamics.
The resulting dataset contains event-level records and structured daily
representations for 911 individuals over 184 days, forming the empirical basis for
the proposed UrbanShare-MoE-PA framework.

\subsection{Real-World Mobility Trajectories and Study Context}

The analysis is based on real-world individual travel trajectory data collected in
Singapore during the COVID-19 pandemic. The dataset records daily activity--travel
behavior for a panel of individuals from 1 March to 31 August 2020, covering four
distinct phases of Singapore's NPIs. The final curated dataset includes
$N=911$ individuals observed over $T=184$ days, yielding 167,624 individual-day
records.

Each individual is associated with key sociodemographic attributes, including age
group, gender, and home location. The trajectory records are linked with detailed
spatial infrastructure, including the road network, public transport system, walking
and cycling networks, POIs, and administrative boundaries. All spatial data are
projected in the SVY21 coordinate system (EPSG:3414), ensuring spatial consistency
across data sources. By using real individual trajectories rather than aggregate
mobility indicators, the dataset enables the reconstruction of daily activity
chains, POI-category exposure contexts, and travel-mode allocation patterns at
the individual level.

\subsection{Semantic Enrichment and Activity Chain Reconstruction}

To transform raw trajectories into behaviorally meaningful representations, we
reconstruct activity chains for each individual as sequences of alternating stay
and travel events. Stay events correspond to stationary periods exceeding five
minutes and are defined by temporal intervals
$[t_{\mathrm{start}}, t_{\mathrm{end}}]$, whereas travel events capture movements
between consecutive stays. Travel distances are computed using the Haversine
formula, and transport modes are inferred based on network and trajectory
characteristics.

The reconstructed activity chains are enriched by integrating external semantic and
spatial data. POIs and areas of interest (AOIs) are derived from OpenStreetMap and
complementary sources such as Foursquare, providing contextual information for
activity inference. The POI system is organized into 12 functional categories,
covering residential, commercial, recreational, institutional, and other urban
activities. This semantic classification enables the transformation of raw movement
records into interpretable daily activity allocations. Administrative boundaries at
multiple spatial scales are further incorporated to support spatial aggregation and
comparative analysis.

Additional contextual layers are introduced to improve behavioral interpretation.
Home locations are used to identify home-based activity, impute missing POI
assignments, and construct distance-from-home features. Demographic attributes are
used to represent heterogeneity across individuals. Policy-related information,
including NPI phase labels and temporal proximity to policy transitions, is
incorporated to explicitly represent exogenous constraints on daily mobility.
Through this multi-source enrichment process, raw trajectories are converted into
structured, semantically meaningful representations of daily urban behavior.

\subsection{Policy Phases and Temporal Encoding}

The study period covers four NPI phases corresponding to Singapore's COVID-19
response, each characterized by different levels of mobility restriction and
activity regulation. Rather than treating these phases only as categorical labels,
we encode them as structured temporal signals to capture both discrete regime shifts
and within-phase behavioral adaptation.

Specifically, each date is associated with a phase label
$p_t \in \{0,1,2,3\}$ and continuous temporal features, including the number of
days since the current phase began and the number of days since the most recent
policy transition. This design allows the model to distinguish abrupt policy
changes from gradual adaptation within a given phase. In addition, a binary
indicator is used to distinguish pre-intervention and intervention periods,
providing a coarse representation of policy activation. These policy features are
later used both as explicit input variables and as the basis for learnable phase
embeddings in the behavioral model.

\subsection{Daily Behavioral Representation and Contextual Features}

Daily behavior is represented using a compositional action vector:
\begin{equation}
a_{it}
=
[t_{it};\,
\mathbf{p}^{c}_{it};\,
\mathbf{p}^{m}_{it}]
\in \mathbb{R}^{1+C+M},
\label{eq:dataset_action}
\end{equation}
where $t_{it}$ denotes the fraction of daily time spent traveling,
$\mathbf{p}^{c}_{it}$ denotes the distribution of non-travel stay time across
POI categories, and $\mathbf{p}^{m}_{it}$ denotes the distribution of travel
time across transport modes, including the unknown-mode channel. This
representation captures the structure of daily time allocation rather than
reducing mobility to a single scalar index. It therefore provides a common
behavioral target for supervised learning, preference alignment, and
alternative-calendar simulation.

To model the determinants of daily behavior, we construct a raw context feature
vector $\mathbf{x}^{\mathrm{ctx}}_{it}$ for each individual--day pair. This vector
integrates day-level contextual variables, individual attributes,
policy-related features, and lagged behavioral history
$\mathbf{h}^{\mathrm{hist}}_{it}$. The day-level contextual variables include
calendar effects such as day of week, observation duration, and temporal
indicators. Individual attributes include age group, gender, and home-based spatial
information. Policy-related features include phase labels, intervention indicators,
days since the current phase began, and temporal distance from policy transitions.
Lagged behavioral features capture short-term persistence and habit formation
through moving averages of recent travel fraction, category shares, and mode
shares.

Policy phase information is intentionally encoded through multiple channels,
including explicit indicators, temporal features, and learnable embeddings in the
model. This design ensures that policy signals remain salient in the
high-dimensional feature space and are not overshadowed by other contextual
variables. Such redundancy is important because behavioral responses in the study
period are strongly shaped by external intervention regimes.

\subsection{Train--Test Split Strategy}

To evaluate model performance under temporally evolving policy conditions, we adopt
a phase-stratified last-week test strategy. For each NPI phase, the final seven
days are held out as the test set, while all preceding observations within the same
phase are used for training. This split evaluates whether the model can reconstruct
recent behavioral patterns within each policy regime after learning from earlier
days of the same phase.

Compared with a random split, this strategy preserves within-phase temporal
ordering while ensuring that all NPI regimes are represented in both training
and test periods. This design supports evaluation of phase-conditioned behavioral
modeling across unseen dates within each observed policy regime.

\section{Methodology}
\label{sec:methodology}

\subsection{UrbanShare-MoE-PA}
\label{sec:model}

\subsubsection{Overview of the Proposed Framework}

\begin{figure}[h]
\centering
\includegraphics[scale=0.5]{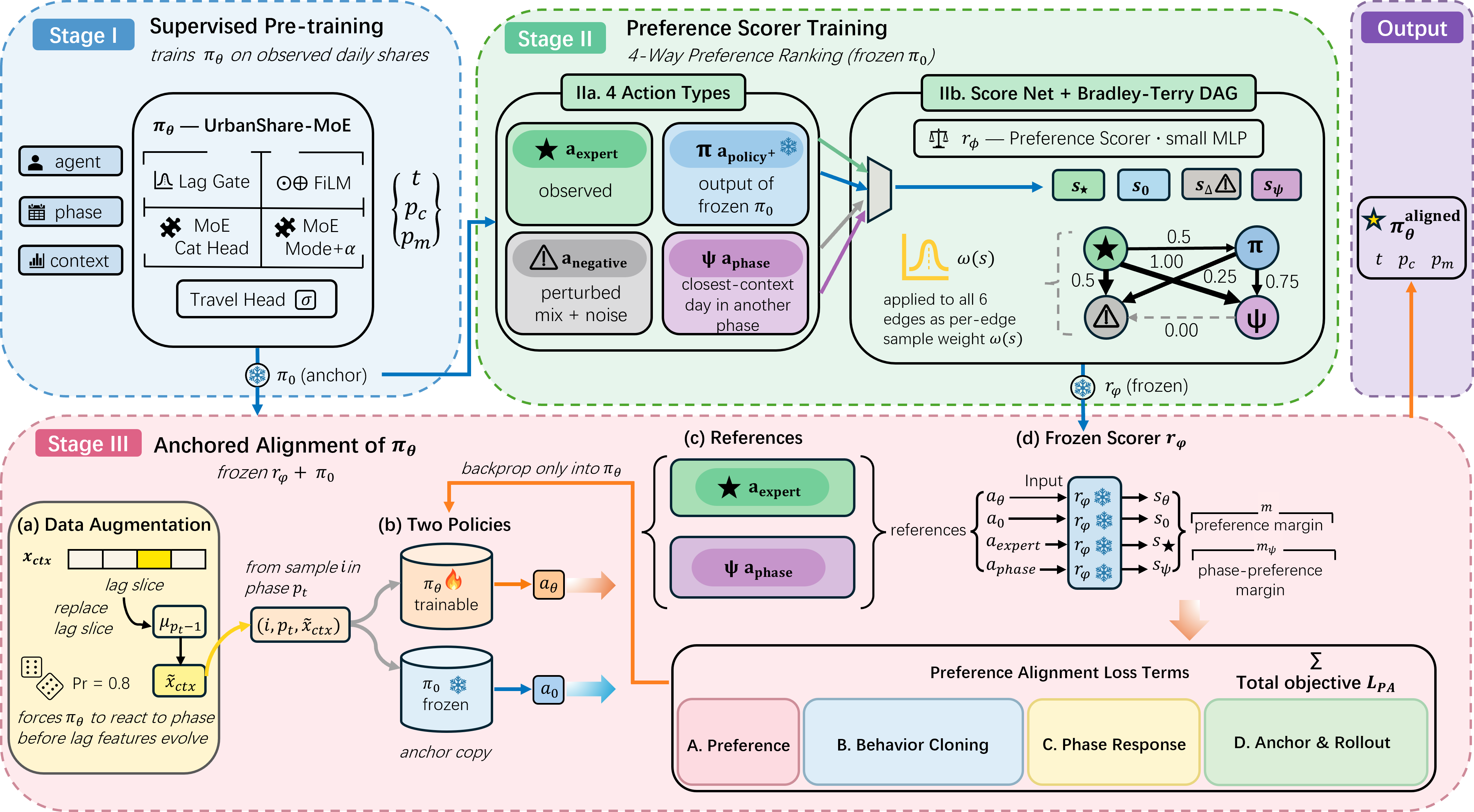}
\caption{UrbanShare-MoE-PA framework.}
\label{FIG:urbanshare-moe-pa}
\end{figure}

UrbanShare-MoE-PA is designed as a behavior-driven scenario simulation
framework that links phase-conditioned daily mobility allocation,
preference-aligned behavioral rollout, and downstream epidemic assessment.
As shown in Figure~\ref{FIG:urbanshare-moe-pa}, the framework contains three
connected stages.

The first stage is supervised pre-training, where the UrbanShare-MoE
behavioral policy $\pi_{\theta}$ is trained on observed daily share
allocations. For each individual--day pair, the model takes agent identity,
NPI phase, and contextual mobility features as inputs, and predicts a
structured daily action
$a_{it}=[t_{it};\mathbf{p}^{c}_{it};\mathbf{p}^{m}_{it}]$, including travel
fraction, POI-category share, and travel-mode share. This stage provides the
baseline behavioral generator and learns the mapping from phase-conditioned
urban context to daily mobility allocation.

The second stage is preference scorer training. A frozen copy of the
supervised policy, denoted as $\pi_{0}$, is used as the anchor policy, while
a preference scorer $r_{\phi}$ is trained to rank candidate actions under the
same agent--phase--context condition. The scorer learns to prefer observed
and phase-consistent behaviors over random or phase-inconsistent
alternatives, with larger weights assigned to samples close to policy
transition boundaries.

The third stage is anchored policy alignment. The trainable policy
$\pi_{\theta}$ is fine-tuned using the frozen preference scorer while being
anchored to the supervised reference policy $\pi_{0}$. This stage encourages
the aligned policy $\pi_{\theta}^{\mathrm{aligned}}$ to generate behaviorally
plausible and phase-consistent actions under both factual and alternative
policy calendars. The aligned policy is then used to recursively simulate
calendar-conditioned daily mobility trajectories, which are subsequently
translated into POI-category-hour and mode-category-hour inputs for
behavior-driven SEIR simulation and scenario comparison.

\subsubsection{Input Features and Structured Baseline}

For each individual--day pair $(i,t)$, we construct a context feature vector
$\mathbf{x}^{\mathrm{ctx}}_{it}$ that summarizes temporal, spatial,
policy-related, recent-behavioral, and demographic information.

Agent identity and policy phase are represented by learnable embeddings, and
the raw context vector $\mathbf{x}^{\mathrm{ctx}}_{it}$ is encoded into a
latent context representation:
\begin{equation}
\mathbf{g}_{it}
=
f_{\mathrm{ctx}}(\mathbf{x}^{\mathrm{ctx}}_{it}).
\label{eq:context_encoder}
\end{equation}

\begin{figure}[h]
\centering
\includegraphics[scale=0.5]{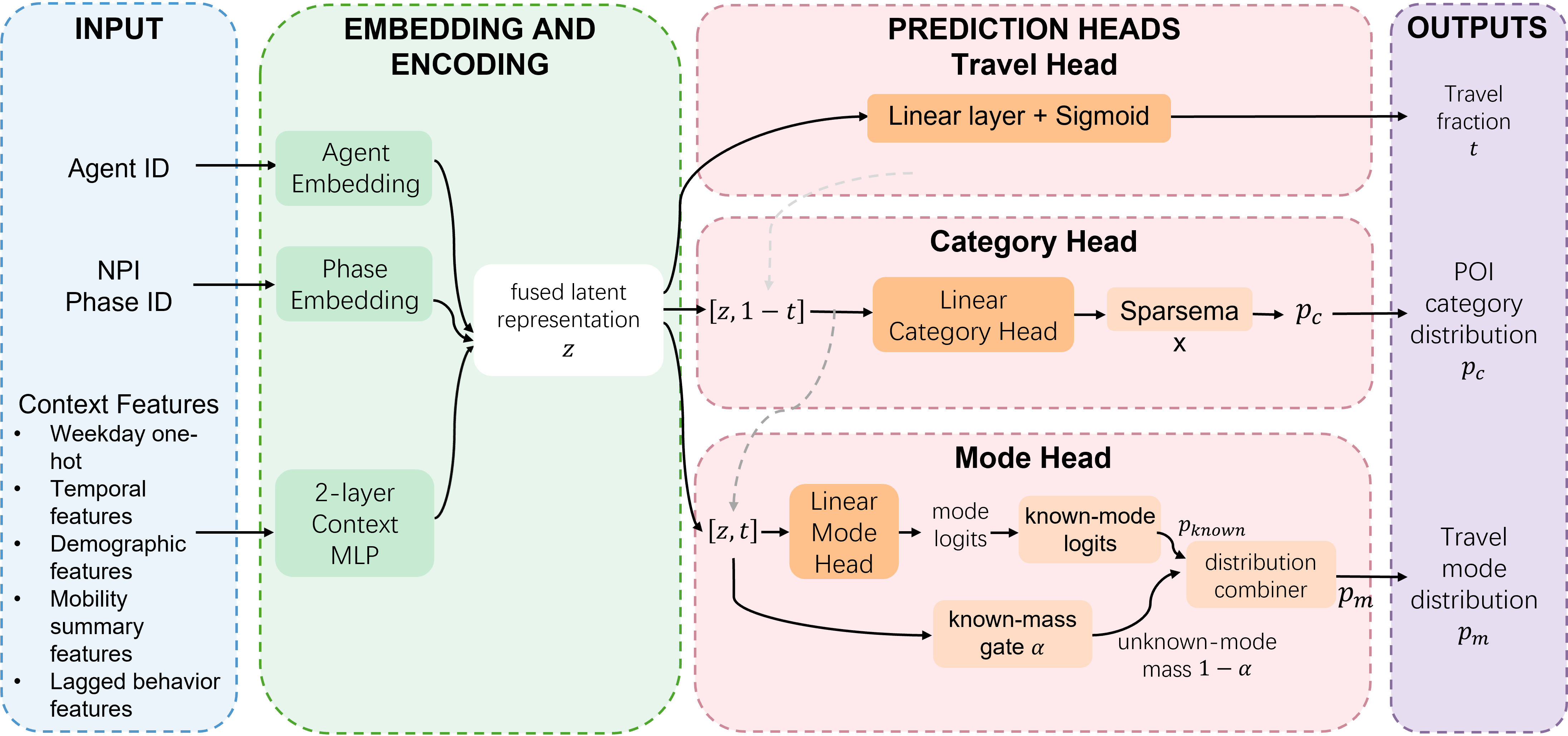}
\caption{UrbanShare baseline.}
\label{FIG:urbanshare}
\end{figure}

\textbf{UrbanShare} is used as the structured baseline, as illustrated in
Figure~\ref{FIG:urbanshare}. It concatenates the agent embedding, phase
embedding, and context encoding into a fused representation, and then passes
it through three prediction heads. The model predicts the travel fraction
$t_{it}$ through a sigmoid head, the POI category share vector
$\mathbf{p}^{c}_{it}$ through a sparsemax projection over the POI categories,
and the travel mode share vector $\mathbf{p}^{m}_{it}$ through an analogous
sparsemax projection over travel modes. For mode prediction, a separate gate
allocates part of the probability mass to an unknown mode, which absorbs
residual uncertainty in mode allocation. Sparsemax
\citep{martins2016softmax} is used to encourage concentrated rather than
fully diffuse daily allocations, matching the empirical fact that most
individuals spend time in only a small number of activity categories and
travel modes on a given day.

UrbanShare uses this fused representation with single-pathway heads and no
policy-conditioned modulation of the context vector, and therefore serves as
the structured baseline of our framework. The following three subsections
introduce phase-gated lag suppression, FiLM-based phase modulation, and
mixture-of-experts routing; combining these three components with the same
hierarchical head structure yields the second progressive model,
UrbanShare-MoE.

\subsubsection{Phase-Gated Lag Suppression}
\label{sec:lag_gate}

Recent-behavior lag features, such as yesterday's home share and the
seven-day rolling mean of travel fraction, are strong predictors of daily
allocations while policy phases remain stable. Under alternative calendars,
however, these lag features preserve the inertia of the historically observed
sequence and can override the current phase indicator, producing behavioral
trajectories that ignore the shifted intervention timing. To attenuate this
failure mode at the input level, UrbanShare-MoE introduces a phase-gated lag
suppression module that gates only the lag component of
$\mathbf{x}^{\mathrm{ctx}}_{it}$ using the current phase embedding:
\begin{equation}
\boldsymbol{\alpha}^{\mathrm{lag}}_{t}
=
\sigma\!\left(\mathbf{W}_{\mathrm{lg}}\mathbf{E}^{p}_{t}\right),
\qquad
\tilde{\mathbf{x}}^{\mathrm{ctx}}_{it,\mathrm{lag}}
=
\mathbf{x}^{\mathrm{ctx}}_{it,\mathrm{lag}}
\odot
\boldsymbol{\alpha}^{\mathrm{lag}}_{t},
\label{eq:lag_gate}
\end{equation}
where $\mathbf{W}_{\mathrm{lg}}$ is a learnable projection from the phase
embedding to the lag-feature dimension, $\sigma(\cdot)$ is the element-wise
sigmoid, and $\odot$ denotes the element-wise product. Non-lag components
(temporal, spatial, and demographic entries) pass through unchanged. For
UrbanShare-MoE, the context encoder in
Equation~\ref{eq:context_encoder} operates on the gated input:
\begin{equation}
\mathbf{g}_{it}
=
f_{\mathrm{ctx}}(\tilde{\mathbf{x}}^{\mathrm{ctx}}_{it}),
\label{eq:context_encoder_gated}
\end{equation}
so that context encoding is applied to a phase-modulated input rather than
to the raw context vector used by the UrbanShare baseline. The gate learns to
attenuate lag influence in phases where recent behavior is expected to
diverge from the current regime, and to preserve it under phases dominated
by autoregressive persistence.

\subsubsection{FiLM-Based Phase Modulation}

The lag-gated context representation $\mathbf{g}_{it}$ from
Equation~\ref{eq:context_encoder_gated} is further modulated by the current
phase before being fused with the agent and phase embeddings. Even after
gating the lag features, simple concatenation of $\mathbf{g}_{it}$ with the
phase embedding may underuse policy information by treating phase as only one
more covariate in a large latent space. To make policy regime changes act as
an explicit conditioning signal on the encoded context itself,
\textbf{UrbanShare-MoE} further applies Feature-wise Linear Modulation (FiLM)
\citep{perez2018film} to $\mathbf{g}_{it}$:
\begin{equation}
\tilde{\mathbf{g}}_{it}
=
\mathbf{g}_{it}
\odot
\left(
1+\tanh(\mathbf{W}_{\gamma}^{\mathrm{film}}\mathbf{E}^{p}_{t})
\right)
+
\mathbf{W}_{\beta}^{\mathrm{film}}\mathbf{E}^{p}_{t},
\label{eq:film}
\end{equation}
where the current phase embedding generates feature-wise scaling and shifting
coefficients. This allows the same contextual state to imply different
mobility responses under different intervention regimes, rather than forcing
phase effects to enter only through static concatenation.

The final fused representation is constructed as
\begin{equation}
\mathbf{z}_{it}
=
[
\mathbf{E}^{a}_{i};
\mathbf{E}^{p}_{t};
\tilde{\mathbf{g}}_{it}
].
\label{eq:fused_representation}
\end{equation}

\subsubsection{Top-$K$ Mixture-of-Experts Heads}

Individuals may follow markedly different allocation rules even under the
same policy regime and contextual conditions. To capture this heterogeneity,
\textbf{UrbanShare-MoE} replaces the single-pathway category and mode heads
with sparse Top-$K$ Mixture-of-Experts (MoE) modules.

UrbanShare-MoE uses a hierarchical factorization consistent with the
behavioral structure. The travel head first predicts the travel fraction
$t_{it}$. The category head then conditions on the remaining non-travel
time, while the mode head conditions on the travel component:
\begin{equation}
\mathbf{z}^{c}_{it}
=
[\mathbf{z}_{it};\mathrm{sg}(1-t_{it})],
\label{eq:z_cat}
\end{equation}
\begin{equation}
\mathbf{z}^{m}_{it}
=
[\mathbf{z}_{it};\mathrm{sg}(t_{it})],
\label{eq:z_mode}
\end{equation}
where $\mathrm{sg}(\cdot)$ denotes the stop-gradient operator. This design
reflects the decomposition that destination-category allocation is
conditional on the available activity time, while mode allocation is
conditional on travel occurrence.

For each head-specific representation $\mathbf{z}^{u}_{it}$, where
$u\in\{c,m\}$, a learnable gate routes each sample to a small subset of
experts:
\begin{equation}
\mathbf{g}^{u}_{it}
=
\mathrm{softmax}\!\left(\mathbf{W}^{u}_{g}\mathbf{z}^{u}_{it}\right).
\label{eq:moe_gate}
\end{equation}
UrbanShare-MoE then activates only the top-$K$ experts and renormalizes
their gate weights:
\begin{equation}
\mathbf{o}^{u}_{it}
=
\sum_{e\in\mathcal{T}^{u}_{it}}
\hat{g}^{u}_{it,e} F^{u}_{e}(\mathbf{z}^{u}_{it}),
\quad
\hat{g}^{u}_{it,e}
=
\frac{g^{u}_{it,e}}
{\sum_{j\in\mathcal{T}^{u}_{it}}g^{u}_{it,j}},
\quad
\mathcal{T}^{u}_{it}
=
\mathrm{TopK}(\mathbf{g}^{u}_{it},K_{\mathrm{top}}),
\label{eq:topk_moe}
\end{equation}
where $F^{u}_{e}(\cdot)$ denotes the $e$-th expert network for head $u$,
$\mathcal{T}^{u}_{it}$ is the selected expert set, and
$\hat{g}^{u}_{it,e}$ denotes the renormalized gate weight. The category head
maps the MoE output to a sparsemax POI-category distribution, while the mode
head produces a sparsemax distribution over observed modes together with an
unknown-mode mass gate. During training, gate noise and a load-balancing
regularizer are used to reduce routing collapse and encourage expert
specialization across behavioral subpopulations.

\subsubsection{Supervised Training with Phase-Sensitivity Regularization}

The supervised objective combines distributional reconstruction losses for
POI-category and mode shares, a travel-fraction loss, hour-level
reconstruction losses, sparsity-preserving penalties, and MoE regularization.
To discourage phase-insensitive behavior, we additionally introduce a
contrastive phase-sensitivity term that penalizes predictions from similar
contexts when their outputs remain too similar across different policy
phases.

The complete supervised training objective is
\begin{equation}
\mathcal{L}_{\mathrm{sup}}
=
\mathcal{L}_{\mathrm{dist}}
+
\lambda_{t}\mathcal{L}_{t}
+
\lambda_{\mathrm{hour}}
\left(
\mathcal{L}_{h^{c}}
+
\mathcal{L}_{h^{m}}
\right)
+
\lambda_{\mathrm{sparse}}\mathcal{L}_{\mathrm{sparse}}
+
\lambda_{\mathrm{aux}}\mathcal{L}_{\mathrm{aux}}
+
\lambda_{\mathrm{ps}}\mathcal{L}_{\mathrm{ps}},
\label{eq:supervised_loss}
\end{equation}
where $\mathcal{L}_{\mathrm{dist}}$ compares predicted and observed share
distributions, $\mathcal{L}_{t}$ supervises travel fraction,
$\mathcal{L}_{h^{c}}$ and $\mathcal{L}_{h^{m}}$ supervise implied
category-hours and mode-hours, $\mathcal{L}_{\mathrm{sparse}}$ preserves
sparse allocations, $\mathcal{L}_{\mathrm{aux}}$ stabilizes MoE routing, and
$\mathcal{L}_{\mathrm{ps}}$ promotes phase responsiveness. Full optimization
hyperparameters are fixed across experiments.

The supervised UrbanShare-MoE model provides a phase-conditioned and
heterogeneity-aware behavioral generator, but its outputs are still
optimized primarily for factual reconstruction. To improve coherence under
shifted policy calendars, we further align the generator using phase-aware
preference ranking.

\subsection{Phase-Aware Preference Alignment}
\label{sec:pa}

While $\mathcal{L}_{\mathrm{ps}}$ regularizes cross-phase distributional
separation during supervised training, it does not directly specify which
generated action is more phase-consistent under alternative policy calendars.
Phase-aware preference alignment addresses this limitation by fine-tuning
the behavioral policy to rank phase-consistent actions above
phase-inconsistent alternatives \citep{song2024preference}. As illustrated in
Figure~\ref{FIG:urbanshare-moe-pa}, this stage introduces a preference
scorer and an anchored alignment procedure to improve response coherence
near policy transition points rather than optimizing same-distribution
reconstruction accuracy alone.

\subsubsection{Action Construction and Preference Scorer}

For each individual--day sample $(i,t)$, the behavioral action
$a_{it}=[t_{it};\mathbf{p}^{c}_{it};\mathbf{p}^{m}_{it}]$ combines travel
fraction, POI-category shares, and mode-category shares. Four action variants
are constructed for preference learning: an observed reference action
$a^{\mathrm{expert}}_{it}$, the current policy action
$a^{\mathrm{policy}}_{it}$, a generic perturbed negative action
$a^{\mathrm{negative}}_{it}$, and a wrong-phase action
$a^{\mathrm{phase}}_{it}$ retrieved from a different policy phase under
similar context. The preference scorer $r_{\phi}(\mathbf{z}_{it},a)$
evaluates each candidate under the current agent, context, and policy-phase
condition.

The preference scorer $r_{\phi}$ is trained with weighted pairwise
Bradley--Terry losses over the six ordered pairs listed in
Table~\ref{tab:bt_pairs}. Each pair $(a\succ b)$ contributes a softplus
ranking term
$w_{ab}\,\omega(s_{it})\,\mathrm{softplus}\!\left(
-\big(r_{\phi}(\mathbf{z}_{it},a)-r_{\phi}(\mathbf{z}_{it},b)\big)
\right)$,
where $w_{ab}$ is the pair-specific weight and $\omega(s_{it})$ is a
transition-proximity weight defined below. Setting the expert vs.\
wrong-phase weight highest concentrates the scorer's capacity on the ranking
dimension that most directly supports alternative-calendar generalization:
distinguishing phase-consistent actions from actions that are locally
plausible but retrieved from a different policy regime. The wrong-phase
vs.\ negative pair is disabled by default to avoid conflating perturbation
noise with phase mismatch.

\begin{table}[h]
\centering
\caption{Weighted pairwise preference relations enforced during scorer
training. Expert vs.\ wrong-phase receives the largest weight, reflecting the
core role of phase consistency in alternative-calendar behavioral simulation.}
\label{tab:bt_pairs}
\small
\setlength{\tabcolsep}{5pt}
\begin{tabular}{lcl}
\toprule
Ordered pair $(a \succ b)$ & Weight $w_{ab}$ & Interpretation \\
\midrule
$a^{\mathrm{expert}} \succ a^{\mathrm{phase}}$    & $1.00$ & expert above wrong-phase (phase rank) \\
$a^{\mathrm{policy}} \succ a^{\mathrm{phase}}$    & $0.75$ & current policy above wrong-phase \\
$a^{\mathrm{expert}} \succ a^{\mathrm{policy}}$   & $0.50$ & expert above current policy \\
$a^{\mathrm{expert}} \succ a^{\mathrm{negative}}$ & $0.50$ & expert above generic negative \\
$a^{\mathrm{policy}} \succ a^{\mathrm{negative}}$ & $0.25$ & current policy above generic negative \\
$a^{\mathrm{phase}}  \succ a^{\mathrm{negative}}$ & $0.00$ & disabled by default \\
\bottomrule
\end{tabular}
\end{table}

The transition-proximity weight $\omega(s_{it})$ upweights samples close to a
recent phase boundary and decays back to unity in the middle of a stable
phase:
\begin{equation}
\omega(s_{it})
=
1 + b\,\exp\!\left(-\frac{d_{it}}{\tau}\right),
\label{eq:omega_s}
\end{equation}
where $d_{it}$ is the number of days since the most recent NPI transition,
and $(b,\tau)$ control the magnitude and decay length of the boundary boost.
In our experiments we use $b=3$ and $\tau=3$ days, so that samples on the
transition day itself carry approximately four times the base weight and
decay back to nominal weight within roughly a week. This concentrates
preference supervision where phase-appropriate adjustment matters most,
without over-fitting the scorer to the smaller number of transition-day
samples.

\subsubsection{Anchored Policy Fine-Tuning}

After the preference scorer is trained, the behavioral policy is aligned
while both the scorer $r_{\phi}$ and the supervised reference policy
$\pi_{0}$ are frozen. Let $\pi_{\theta}$ denote the trainable policy,
initialised from $\pi_{0}$.

\paragraph{Distribution-shift augmentation.}
Under the factual calendar, lagged behavioral features and phase indicators
co-evolve continuously, so the trained policy rarely encounters a phase--lag
mismatch. Under alternative calendars, however, a shifted phase sequence
assigns a new phase to a sample whose lag features still reflect the
historical phase, and the policy must learn to condition on the new phase
rather than defer to autoregressive inertia. To expose $\pi_{\theta}$ to
this input distribution during training, we apply an augmentation prior to the alignment forward pass. For every training sample
with a valid preceding phase, the lag component of the context vector is
replaced, with probability $p_{\mathrm{aug}}=0.8$, by the population-mean
lag prototype $\boldsymbol{\mu}^{\mathrm{lag}}_{p_{t}-1}$, obtained by averaging the lag features over all individual–day samples assigned to the preceding phase:
\begin{equation}
\mathbf{x}^{\mathrm{ctx}}_{it,\mathrm{lag}}
\;\leftarrow\;
\boldsymbol{\mu}^{\mathrm{lag}}_{p_{t}-1}
\quad
\text{with probability } p_{\mathrm{aug}}.
\label{eq:cf_augmentation}
\end{equation}
This substitution mimics the input distribution that arises under a
shifted-calendar rollout, where the current phase has changed but the lag
window has not yet caught up. Without this augmentation, the aligned policy
tends to reproduce the observed phase--lag correlation and to under-respond
at alternative-calendar transition boundaries.

\paragraph{Adaptive preference margins.}
The scorer-driven gradient into $\pi_{\theta}$ is not applied uniformly
across samples: it is modulated by two detached, scorer-derived confidence
margins. Let
$s^{\ast}_{it}=r_{\phi}(\mathbf{z}_{it},a^{\mathrm{expert}}_{it})$,
$s^{0}_{it}=r_{\phi}(\mathbf{z}_{it},a^{\mathrm{policy},0}_{it})$,
$s^{\psi}_{it}=r_{\phi}(\mathbf{z}_{it},a^{\mathrm{phase}}_{it})$, and
$s^{\theta}_{it}=r_{\phi}(\mathbf{z}_{it},a^{\theta}_{it})$
denote the frozen scorer's evaluations of the expert, anchor, wrong-phase,
and current trainable actions. We define
\begin{equation}
m_{it}
=
\mathrm{sg}\!\big(\sigma(s^{\ast}_{it}-s^{0}_{it})\big),
\qquad
m^{\psi}_{it}
=
\mathrm{sg}\!\big(\sigma(s^{\ast}_{it}-s^{\psi}_{it})\big),
\label{eq:pa_margins}
\end{equation}
where $\mathrm{sg}(\cdot)$ denotes the stop-gradient operator. The
\emph{preference margin} $m_{it}\in(0,1)$ measures how confidently the
scorer prefers the observed action over the anchor policy output, and
modulates the preference term $\mathcal{L}_{\mathrm{pref}}$ so that
preference-driven updates are strong only where the scorer separates the
two actions cleanly. The \emph{phase-preference margin} $m^{\psi}_{it}$
measures how confidently the scorer distinguishes the observed action from
the wrong-phase retrieval, and modulates both the scorer-based phase term
$\mathcal{L}_{\mathrm{phase\_pref}}$ and the distributional phase term
$\mathcal{L}_{\mathrm{phase\_wrong}}$ defined below. Because both margins
are detached, they contribute only as per-sample confidence weights and do
not backpropagate through the scorer.

\paragraph{Alignment objective.}
The trainable policy is optimised with a composite objective grouped into
four blocks: (A) scorer-based preference, (B) behavior cloning,
(C) phase response, and (D) anchor and rollout:
\begin{equation}
\mathcal{L}_{\mathrm{PA}}
=
\underbrace{\lambda_{\mathrm{pref}}\mathcal{L}_{\mathrm{pref}}
+\lambda_{\mathrm{ppf}}\mathcal{L}_{\mathrm{phase\_pref}}}_{\text{(A) preference}}
+
\underbrace{\lambda_{\mathrm{BC}}\mathcal{L}_{\mathrm{BC}}
+\lambda_{\mathrm{zero}}\mathcal{L}_{\mathrm{zero}}}_{\text{(B) behavior cloning}}
+
\underbrace{\lambda_{\mathrm{pp}}\mathcal{L}_{\mathrm{post\_phase}}
+\lambda_{\mathrm{pw}}\mathcal{L}_{\mathrm{phase\_wrong}}}_{\text{(C) phase response}}
+
\underbrace{\lambda_{\mathrm{anc}}\mathcal{L}_{\mathrm{anchor}}
+\lambda_{\mathrm{roll}}\mathcal{L}_{\mathrm{rollout}}}_{\text{(D) anchor \& rollout}},
\label{eq:pa_total}
\end{equation}
where the terms are defined as follows. Block~(A) applies scorer-derived
preference signals gated by the margins in
Equation~\ref{eq:pa_margins}:
$\mathcal{L}_{\mathrm{pref}}=-\,m_{it}\,s^{\theta}_{it}$ pulls the scorer
value of the trainable policy upward, and
$\mathcal{L}_{\mathrm{phase\_pref}}
=m^{\psi}_{it}\,\mathrm{softplus}\!\big(-(s^{\theta}_{it}-s^{\psi}_{it})\big)$
enforces separation from the wrong-phase action in scorer space.
Block~(B) aggregates KL and MSE reconstruction terms over POI-category,
mode-category, travel-fraction, and category- or mode-hour allocations
($\mathcal{L}_{\mathrm{BC}}$), together with sparsity penalties on the
category and mode simplices ($\mathcal{L}_{\mathrm{zero}}$). These terms
prevent the preference gradients from degrading same-distribution fit.
Block~(C) contains two hinge-style penalties on the aligned policy's
distributional output:
$\mathcal{L}_{\mathrm{post\_phase}}$ requires the aligned distributions to
diverge from the preceding-phase reference by at least a margin
$\mu^{\mathrm{prev}}$, and
$\mathcal{L}_{\mathrm{phase\_wrong}}
=m^{\psi}_{it}\,\mathrm{ReLU}\!\big(\mu^{\psi}
-\mathrm{KL}(\hat{p}^{\theta}_{it}\,\|\,\hat{p}^{\psi}_{it})\big)$
requires them to diverge from the wrong-phase retrieval by at least a
margin $\mu^{\psi}$, weighted again by the phase-preference margin.
Block~(D) regularises the aligned policy toward the supervised reference:
$\mathcal{L}_{\mathrm{anchor}}$ combines KL and MSE anchors on POI-category,
mode-category, and travel-fraction outputs against the frozen $\pi_{0}$
outputs, and $\mathcal{L}_{\mathrm{rollout}}$ enforces one-step temporal
consistency between successively generated actions along the rollout
trajectory. Anchoring keeps the aligned policy behaviorally close to the
supervised reference on the observed distribution while allowing the
preference and phase-response terms to reshape its behavior at policy
transitions. All $\lambda$ hyperparameters are fixed across factual and
alternative-calendar runs.

In the reported experiments, a single preference-aligned checkpoint is used
for factual reconstruction, alternative-calendar rollout, and downstream
SEIR evaluation. This choice keeps the evaluation tied to one behavioral
policy rather than mixing separate checkpoints for factual and
alternative-calendar analyses. The resulting aligned policy is denoted as
$\pi_{\theta}^{\mathrm{aligned}}$ and is used for all UrbanShare-MoE-PA
trajectory generation.

\subsection{Calendar-Conditioned Policy Scenario Simulation}
\label{sec:cf}

Given the preference-aligned behavioral policy
$\pi_{\theta}^{\mathrm{aligned}}$, alternative-calendar trajectories are
generated by replacing the factual phase sequence $\{p_t\}_{t=1}^{T}$ with
an alternative calendar $\{p^{\mathrm{cf}}_t\}_{t=1}^{T}$ while holding
individual attributes and non-policy exogenous contextual variables fixed.
In addition to the phase label, the alternative calendar replaces
calendar-relative policy features such as days-since-phase and
days-since-NPI. In this study, we compare four alternative policy calendars
that vary the timing or duration of Singapore's lockdown phase. The four
calendars are summarized in Table~\ref{tab:policy_calendars}.

\begin{table}[t]
\centering
\caption{Alternative policy calendars.}
\label{tab:policy_calendars}
\small
\setlength{\tabcolsep}{4pt}
\begin{tabular}{p{2.4cm}p{2.6cm}p{3.3cm}p{3.1cm}p{2.7cm}}
\toprule
Scenario & Modification & Lockdown & Phase 1 & Phase 2 \\
\midrule
Early lockdown
& Start $-7$ days
& 2020-03-31--2020-05-25 (56 days)
& 2020-05-26--2020-06-14 (20 days)
& 2020-06-15--2020-08-31 \\
Late lockdown
& Start $+7$ days
& 2020-04-14--2020-06-08 (56 days)
& 2020-06-09--2020-06-28 (20 days)
& 2020-06-29--2020-08-31 \\
Short lockdown
& Duration $-14$ days
& 2020-04-07--2020-05-18 (42 days)
& 2020-05-19--2020-06-07 (20 days)
& 2020-06-08--2020-08-31 \\
Long lockdown
& Duration $+14$ days
& 2020-04-07--2020-06-15 (70 days)
& 2020-06-16--2020-07-05 (20 days)
& 2020-07-06--2020-08-31 \\
\bottomrule
\end{tabular}
\end{table}

Alternative-calendar actions are generated recursively under the modified
policy calendar. On each date, the model receives the alternative calendar
features together with lagged behavioral history
$\mathbf{h}^{\mathrm{hist,cf}}_{it}$, which is updated from previously
generated scenario actions. This recursive update allows prior simulated
actions to affect subsequent behavioral predictions through the
lagged-history component of the state.

Because the full sequence of phase-relative timing features is replaced
under a shifted policy calendar, the simulated trajectory can differ from
the factual trajectory before the historical lockdown date on the absolute
time axis. These pre-boundary differences arise from the modified calendar
conditioning and the recursively updated behavioral history.

In the reported alternative-calendar simulations, the rollout is not
hard-anchored to observed pre-divergence actions, no same-phase
observed-action blending is applied, and no hand-coded POI-category
constraint is imposed. The resulting alternative-calendar daily shares
represent a calendar-conditioned full-trajectory re-simulation driven by
the learned behavioral policy itself and are then aggregated to
population-level time series for the downstream epidemic simulator.

\subsection{Behavior-Driven SEIR Dynamics}
\label{sec:seir}

Factual and alternative-calendar behavioral trajectories are propagated into
an individual-level SEIR compartmental model with states $\{S,E,I,R\}$,
latent period $L=3$ days, and infectious period $D=7$ days. Let
$X_{it}\in\{S,E,I,R\}$ denote the epidemic state of individual $i$ on day
$t$. The daily infection hazard for susceptible individual $i$ is driven by
two behavioral exposure channels: time spent in POI categories and time
spent in travel-mode categories. Let $\tilde{p}_t$ denote the policy phase
used in a simulation run, with $\tilde{p}_t=p_t$ under the factual calendar
and $\tilde{p}_t=p^{\mathrm{cf}}_t$ under an alternative calendar.

For each POI category $c$, the infectious share of category-hours is
\begin{equation}
\rho^{\mathrm{cat}}_{ct}
=
\frac{
\sum_{i':X_{i't}=I} h^{\mathrm{cat}}_{i'ct}
}{
\sum_{i''} h^{\mathrm{cat}}_{i''ct}+\varepsilon
},
\label{eq:poi_prevalence}
\end{equation}
and, analogously, the infectious share of mode-category hours is
\begin{equation}
\rho^{\mathrm{mode}}_{mt}
=
\frac{
\sum_{i':X_{i't}=I} h^{\mathrm{mode}}_{i'mt}
}{
\sum_{i''} h^{\mathrm{mode}}_{i''mt}+\varepsilon
}.
\label{eq:mode_prevalence}
\end{equation}
Here $h^{\mathrm{cat}}_{ict}$ and $h^{\mathrm{mode}}_{imt}$ denote the
24-hour-scaled category and mode hours generated by the behavioral model,
and $\varepsilon$ avoids division by zero.

The daily hazard combines these prevalence terms with category-specific
contact multipliers, phase-level transmission, and behavioral crowding:
\begin{equation}
\lambda_{it}
=
\omega_t
\beta_{\tilde{p}_t}
\left[
\sum_c
\frac{h^{\mathrm{cat}}_{ict}}{24}
r_{c,\tilde{p}_t}
g^{\mathrm{cat}}_{ct}
\rho^{\mathrm{cat}}_{ct}
+
\sum_m
\frac{h^{\mathrm{mode}}_{imt}}{24}
\nu_m
g^{\mathrm{mode}}_{mt}
\rho^{\mathrm{mode}}_{mt}
\right]
\delta^{\mathrm{gender}}_i
\delta^{\mathrm{age}}_i,
\label{eq:daily_hazard}
\end{equation}
where $\beta_{\tilde{p}_t}$ is the calibrated phase-specific transmission
coefficient, $r_{c,\tilde{p}_t}$ and $\nu_m$ are POI-category and
mode-category contact multipliers, and $\delta^{\mathrm{gender}}_i$ and
$\delta^{\mathrm{age}}_i$ are demographic susceptibility multipliers. The
factor $\omega_t$ captures population-level home protection and decreases
transmission when the mean home share is high. The crowding factors
$g^{\mathrm{cat}}_{ct}$ and $g^{\mathrm{mode}}_{mt}$ compare the total
category or mode presence under a scenario with the observed factual
presence on the same date:
\begin{equation}
g^{\mathrm{cat}}_{ct}
=
\left(
\frac{H^{\mathrm{cat}}_{ct}+\varepsilon}
{H^{\mathrm{cat,obs}}_{ct}+\varepsilon}
\right)
^{\xi},
\qquad
g^{\mathrm{mode}}_{mt}
=
\left(
\frac{H^{\mathrm{mode}}_{mt}+\varepsilon}
{H^{\mathrm{mode,obs}}_{mt}+\varepsilon}
\right)
^{\xi}.
\label{eq:crowding}
\end{equation}
The main analysis uses $\xi=1$, so reductions or increases in scenario-level
presence proportionally attenuate or amplify contact intensity relative to
the observed factual reference.

For susceptible individuals, the daily transition probability from
susceptible to exposed is
\begin{equation}
P(X_{i,t+1}=E \mid X_{it}=S)
=
1-
\left(1-\pi_{\mathrm{imp}}\right)\exp(-\lambda_{it}),
\label{eq:se_transition}
\end{equation}
where $\pi_{\mathrm{imp}}$ is a small, fixed background infection
probability. This background term is held constant across dates, policies,
and behavior models; we do not use policy-specific importation multipliers
in the reported experiments. Compartment transitions are updated
synchronously at the end of each simulated day: exposed individuals become
infectious after $L$ days, and infectious individuals recover after $D$
days. Initial seeds are placed directly into the infectious compartment at
the start of the simulation.

Because the simulator operates on a 911-agent sample rather than on the
full Singapore population, epidemic parameters are calibrated with a
reproducible search procedure against normalized trend summaries of
Singapore's reported first-wave epidemic curves
\citep{moh2020pastupdates,cda2025covid} rather than against raw national
counts. Reported cases lag latent infections, so simulated new exposures
are aligned to the 7-day-smoothed reported daily case curve with a fixed
21-day lead. The calibration objective combines four shape-level criteria:
first-wave peak timing, lag-adjusted incidence correlation and NRMSE,
phase-level exposure shares across pre-NPI, lockdown, Phase~1, and Phase~2,
and the normalized cumulative-case growth trajectory. The search tunes the
global transmission scale, the population-mean home-contact attenuation
factor, the fixed background infection probability, and phase-specific
transmission multipliers. The resulting calibrated epidemic parameter set is
held fixed for all factual and alternative-calendar runs so that scenario
differences are evaluated under a common epidemic parameterization rather
than under scenario-specific epidemic re-tuning. The resulting epidemic
outputs are therefore interpreted as model-based scenario comparisons on
the calibrated 911-agent scale rather than as direct national case-count
estimates.

\subsection{Evaluation Protocol}
\label{sec:eval}

We evaluate the behavioral-policy family across three levels: behavioral
fidelity under the factual policy sequence, calendar-conditioned scenario
comparison, and downstream epidemic and activity summaries. Factual
behavioral reconstruction and alternative-calendar figures compare
UrbanShare, UrbanShare-MoE, and UrbanShare-MoE-PA. The policy ranking and
epidemic--activity trade-off analysis are reported for UrbanShare-MoE-PA as
the primary specification, while the other two models provide baselines for
interpreting how behavior-generator quality affects downstream simulations.

\textbf{Behavioral fidelity under the factual policy sequence.}
We evaluate held-out reconstruction and full-horizon factual simulation
against the same observed population-mean behavioral series, so that
held-out predictive accuracy and full-horizon factual fidelity are reported
on a common behavioral scale. For each date $t$ and behavioral series $s$,
let $\bar{h}_{ts}$ and $\hat{\bar{h}}_{ts}$ denote the observed and
simulated population-mean daily hours, respectively:
\begin{equation}
\bar{h}_{ts}
=
\frac{1}{N_t}\sum_i h_{its},
\qquad
\hat{\bar{h}}_{ts}
=
\frac{1}{N_t}\sum_i \hat{h}_{its},
\end{equation}
where $N_t$ is the number of agents observed on date $t$. We report MAE,
RMSE, absolute bias, and Pearson correlation for 12 POI-category-hour
series and 5 observed travel-mode-hour series, macro-averaged within each
family and excluding the unknown-mode channel from the mode-family summary.
MAE and RMSE capture average reconstruction error, absolute bias measures
the magnitude of systematic mean over- or under-prediction after averaging
signed errors over time, and correlation captures temporal co-movement.

Held-out reconstruction uses the last seven days of each NPI phase, giving
28 held-out dates in total, whereas full-horizon factual fidelity uses the
full 184-day factual sequence. We further evaluate the held-out set at the
agent-day distributional level. Specifically, we report weighted KL
divergence for POI-category distributions, for mode-category distributions
over the five observed modes, and for mode distributions including the
unknown-mode channel; weighted mean-squared error for travel fraction and
unknown-mode hours; and $R^2$ for reconstructed category-hours and
mode-hours. Agent-days are weighted by observed available hours, and
mode-distribution terms are additionally weighted by observed travel hours
so that days with negligible travel do not dominate the mode evaluation.

\textbf{Policy scenario comparison.}
We compare the original policy calendar with four alternative lockdown
calendars: early lockdown, late lockdown, short lockdown, and long lockdown.
Early and late lockdown shift the intervention start date by seven days
while keeping the lockdown duration fixed at 56 days. Short and long
lockdown keep the historical start date of 7 April 2020 but change the
lockdown duration to 42 and 70 days, respectively.
Table~\ref{tab:policy_calendars} lists the resulting phase calendars. For
each scenario $q$, the model generates alternative-calendar population-mean
daily-hour trajectories $\hat{\bar{h}}^{(q)}_{ts}$, which are compared with
the factual simulation $\hat{\bar{h}}^{\mathrm{fact}}_{ts}$ over the
post-divergence dates. Because the alternative-calendar trajectories are
not historically observed, the comparison focuses on the direction,
temporal coherence, and differentiation of the induced behavioral changes
across the four policy calendars before examining their downstream epidemic
and activity implications. Replacing the full calendar-conditioning
sequence and recursively updating lagged behavior can also produce modest
pre-boundary deviations on the absolute date axis.

\textbf{Epidemic outcomes.}
The behavior-driven SEIR model is run under each NPI scenario using the
predicted daily behavioral trajectories. Following
\citet{liu2022modelling}, epidemic outcomes are computed from active
infections $I_t$, newly exposed individuals $\mathrm{new}E_t$, and the
final recovered count $R_T$. We report peak infectious prevalence
($\max_t I_t$), the peak 7-day-smoothed number of new exposures,
Final~$R$ on the 911-agent scale, Area($I$), Area(new~$E$), and epidemic
duration, measured as the number of days with $I_t>1$. Peak and area
metrics are treated as the primary burden summaries; duration is reported
as a supplementary tail-sensitive statistic.

\textbf{Sector-weighted urban activity.}
The sector-weighted urban activity index is computed directly from the
behavior layer. For policy trade-off comparisons, the original policy
sequence and each alternative calendar are evaluated on the same
UrbanShare-MoE-PA behavioral scale. Let
$\bar{h}^{(q),\mathrm{cat}}_{tc}$ denote the population-mean daily stay
hours in POI category $c$ on day $t$ under scenario $q$, and let
$\bar{h}^{(q),\mathrm{mode}}_{tm}$ denote the corresponding population-mean
daily travel hours assigned to mode category $m$. The sector-weighted
activity index is measured as a weighted sum of POI-category stay hours and
mobility-related travel hours:
\begin{equation}
O(q)
=
\sum_{t=1}^{T}
\left[
\sum_{c=1}^{C}
\psi_c \cdot \bar{h}^{(q),\mathrm{cat}}_{tc}
+
\psi_{\mathrm{tr}}
\sum_{m=1}^{M}
\bar{h}^{(q),\mathrm{mode}}_{tm}
\right],
\label{eq:econ_output}
\end{equation}
where $\psi_c\in[0,1]$ is the sector weight for POI category $c$, and
$\psi_{\mathrm{tr}}$ is the transportation-sector weight applied to
travel-mode hours. The POI term captures destination-based activity time,
while the travel term captures time spent in mobility itself, including the
unknown-mode channel. Thus, the activity index changes only through changes
in daily behavioral allocations; the SEIR state variables do not enter this
calculation.

The GVA-based sector weights are constructed by mapping POI categories to
the closest Singapore SSIC industry groups and normalizing their official
2020 GVA shares \citep{singstat2021}. Specifically, let $g_c$ denote the
GVA share (\%) of the industry mapped to POI category $c$. The normalized
sector weight is
\begin{equation}
\psi_c
=
\frac{g_c}{\max_{c'}g_{c'}},
\label{eq:econ_weight}
\end{equation}
where the maximum is taken over all non-home categories. The Business \&
professional services category, combining Finance \& Insurance (14.9\%) and
Professional Services (6.1\%), has the highest GVA share among the mapped
consumer-facing service industries and therefore receives $\psi_c=1.0$. All
other weights are computed as the ratio of their GVA share to 21.0\% and
rounded to two decimal places. Home activity is assigned $\psi_c=0$ because
it represents non-market household activity. The travel-mode term uses the
same transportation-sector weight as the travel and transportation POI
category, $\psi_{\mathrm{tr}}=0.31$. Table~\ref{tab:econ_weights} reports
the full mapping and computed weights.

\begin{table}[h]
\centering
\caption{GVA-based sector weights.}
\label{tab:econ_weights}
\begin{tabular}{p{4.2cm}p{5.0cm}ccc}
\toprule
POI Category & Mapped Industry (SSIC 2020) & $g_c$ (\%) & $g_c/21.0$ & $\psi_c$ \\
\midrule
Business \& professional services
    & Finance \& insurance + Professional services
    & 21.0 & 1.000 & 1.00 \\
Travel \& transportation
    & Transportation \& storage
    & 6.5 & 0.310 & 0.31 \\
Education
    & Education services
    & 3.1 & 0.148 & 0.15 \\
Health \& medicine
    & Health \& social services
    & 2.8 & 0.133 & 0.13 \\
Community \& government
    & Public administration \& defence
    & 2.8 & 0.133 & 0.13 \\
Retail
    & Retail trade
    & 1.3 & 0.062 & 0.06 \\
Landmarks \& outdoors
    & Other services
    & 1.3 & 0.062 & 0.06 \\
Event
    & Other services
    & 1.3 & 0.062 & 0.06 \\
Dining \& drinking
    & Food \& beverage services
    & 1.1 & 0.052 & 0.05 \\
Arts \& entertainment
    & Arts, entertainment \& recreation
    & 0.7 & 0.033 & 0.03 \\
Sports \& recreation
    & Arts, entertainment \& recreation
    & 0.7 & 0.033 & 0.03 \\
Home
    & Non-market household activity
    & 0.0 & --- & 0.00 \\
\bottomrule
\end{tabular}
\end{table}

\section{Results}
\label{sec:results}
\subsection{Behavioral Fidelity under the Factual Policy Sequence}
\label{sec:results_factual}
Behavioral fidelity under the factual calendar is evaluated at three
complementary resolutions: held-out macro-averaged reconstruction,
held-out agent-day distributional fidelity, and full-horizon population-level
temporal fidelity.

Table~\ref{tab:behavioral_fidelity} summarizes the macro results. On the
held-out test dates (Panel A), UrbanShare-MoE sharply improves POI-category
reconstruction relative to UrbanShare, reducing MAE from 0.101 to 0.049 and
RMSE from 0.122 to 0.067. UrbanShare-MoE-PA remains nearly as strong on POIs
and gives the best travel-mode error profile, with
MAE~=~0.009, RMSE~=~0.011, and Abs.\ Bias~=~0.002. The same pattern persists
over the full 184-day factual sequence (Panel B): UrbanShare-MoE is strongest
for POI categories, whereas UrbanShare-MoE-PA achieves the lowest travel-mode
errors. This pattern is consistent with the intended roles of the model
components: the MoE variant shows the largest gains in POI reconstruction,
whereas the aligned variant gives the lowest travel-mode error metrics.
\begin{table}[h]
\centering
\caption{Factual behavioral fidelity.}
\label{tab:behavioral_fidelity}
\begin{tabular}{llcccc}
\toprule
\multicolumn{6}{l}{\textbf{Panel A. Held-out reconstruction (test dates)}} \\
\midrule
Model & Output & MAE (h/day) & RMSE (h/day) & Abs.\ Bias (h/day) & Corr. \\
\midrule
UrbanShare
    & POI categories & 0.101 & 0.122 & 0.089 & 0.815 \\
UrbanShare-MoE
    & POI categories & \textbf{0.049} & \textbf{0.067} & \textbf{0.029} & 0.899 \\
UrbanShare-MoE-PA
    & POI categories & 0.053 & 0.069 & 0.035 & \textbf{0.900} \\
\midrule
UrbanShare
    & Travel modes & 0.022 & 0.024 & 0.021 & \textbf{0.933} \\
UrbanShare-MoE
    & Travel modes & 0.013 & 0.015 & 0.010 & 0.914 \\
UrbanShare-MoE-PA
    & Travel modes & \textbf{0.009} & \textbf{0.011} & \textbf{0.002} & 0.921 \\
\midrule
\multicolumn{6}{l}{\textbf{Panel B. Full-horizon factual simulation fidelity (184 days)}} \\
\midrule
UrbanShare
    & POI categories & 0.091 & 0.109 & 0.082 & 0.912 \\
UrbanShare-MoE
    & POI categories & \textbf{0.033} & \textbf{0.045} & \textbf{0.016} & \textbf{0.957} \\
UrbanShare-MoE-PA
    & POI categories & 0.037 & 0.049 & 0.023 & \textbf{0.957} \\
\midrule
UrbanShare
    & Travel modes & 0.021 & 0.024 & 0.021 & 0.954 \\
UrbanShare-MoE
    & Travel modes & 0.013 & 0.015 & 0.011 & 0.958 \\
UrbanShare-MoE-PA
    & Travel modes & \textbf{0.008} & \textbf{0.010} & \textbf{0.003} & \textbf{0.959} \\
\bottomrule
\end{tabular}
\end{table}

\begin{table}[h]
\centering
\scriptsize
\caption{Held-out distributional fidelity.}
\label{tab:distributional_fidelity}
\resizebox{\linewidth}{!}{%
\begin{tabular}{lccccccc}
\toprule
Model & $\mathrm{wKL}_{\mathrm{cat}}$ & $\mathrm{wKL}_{\mathrm{mode}}$ & $\mathrm{wKL}_{\mathrm{mode,all}}$ & $\mathrm{wMSE}_{\mathrm{travel}}$ & $\mathrm{wMSE}_{\mathrm{unk\text{-}h}}$ & $R^2_{\mathrm{cat\text{-}hours}}$ & $R^2_{\mathrm{mode\text{-}hours}}$ \\
\midrule
UrbanShare & 0.786 & \textbf{0.565} & \textbf{0.481} & 0.0046 & 1.725 & 0.289 & 0.356 \\
UrbanShare-MoE & \textbf{0.376} & 0.736 & 0.549 & 0.0046 & 1.724 & 0.500 & 0.350 \\
UrbanShare-MoE-PA & 0.380 & 0.681 & 0.522 & \textbf{0.0045} & \textbf{1.723} & \textbf{0.501} & \textbf{0.365} \\
\bottomrule
\end{tabular}%
}
\end{table}

Table~\ref{tab:distributional_fidelity} evaluates the same test dates at the
agent-day distributional level. UrbanShare-MoE roughly halves
$\mathrm{wKL}_{\mathrm{cat}}$ and raises $R^2_{\mathrm{cat\text{-}hours}}$
from 0.289 to 0.500, showing a much better recovery of POI-category
distributions and implied hours. For travel modes, UrbanShare-MoE-PA gives the
best hour-space calibration, with the lowest
$\mathrm{wMSE}_{\mathrm{travel}}$, the lowest
$\mathrm{wMSE}_{\mathrm{unk\text{-}h}}$, and the highest
$R^2_{\mathrm{mode\text{-}hours}}$. UrbanShare, however, retains the lowest
mode-distribution KL values, showing that distributional similarity and
hour-scale calibration do not rank the models identically. Overall, the MoE
variants provide the largest gains for POI reconstruction, while the aligned
variant gives the strongest travel-mode hour calibration.

Importantly, these gains are produced by an agent-level behavioral engine
rather than by direct fitting of aggregate curves. Each model predicts a full
daily allocation vector for every individual agent, and the population-level
figures are obtained only after aggregating those individual trajectories. This
design retains individual-level variation rather than fitting the aggregate
curves directly. Appendix
Figures~\ref{fig:rep_agent_factual_poi_profiles_a},
\ref{fig:rep_agent_factual_poi_profiles_b}, and
\ref{fig:rep_agent_factual_mode_profiles}
illustrate this point for a representative agent with relatively balanced time
allocation across activities and modes.

\begin{figure*}[t]
\centering
\includegraphics[width=0.78\linewidth]{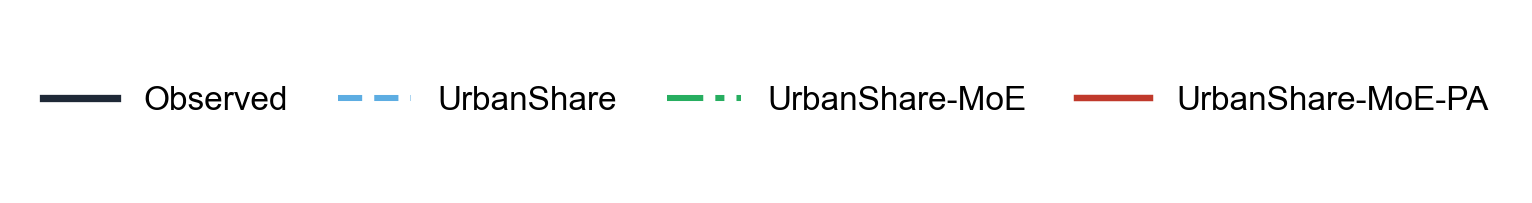}

\medskip

\includegraphics[width=0.48\textwidth]{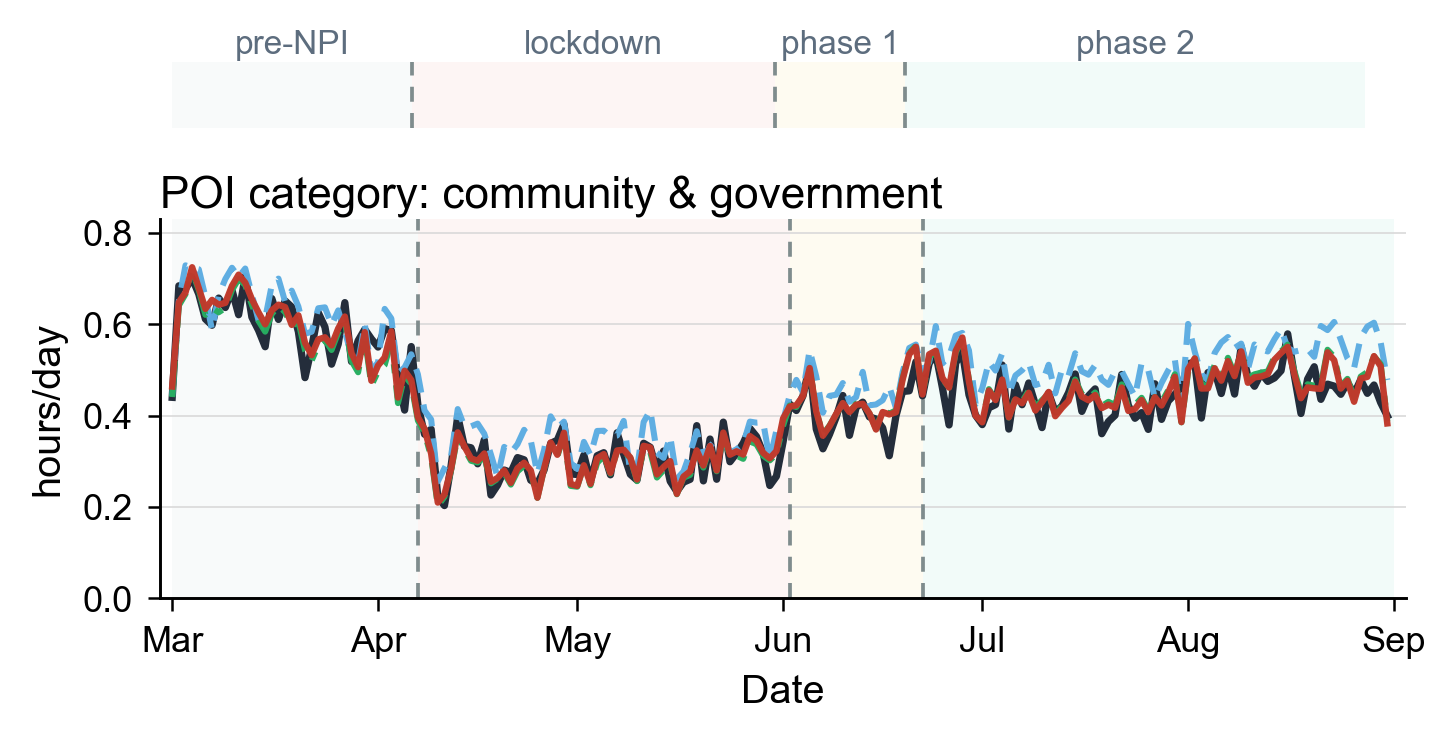}
\hfill
\includegraphics[width=0.48\textwidth]{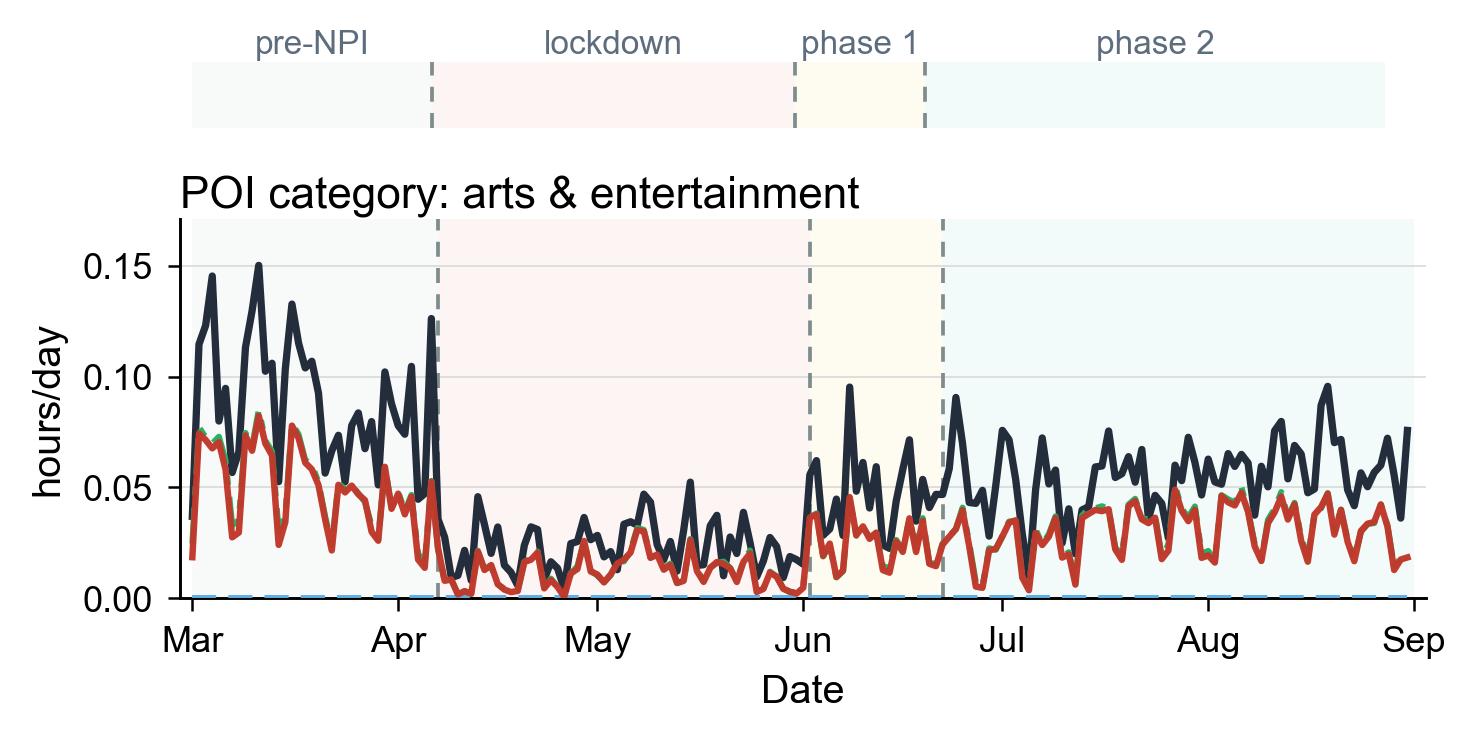}

\medskip

\includegraphics[width=0.48\textwidth]{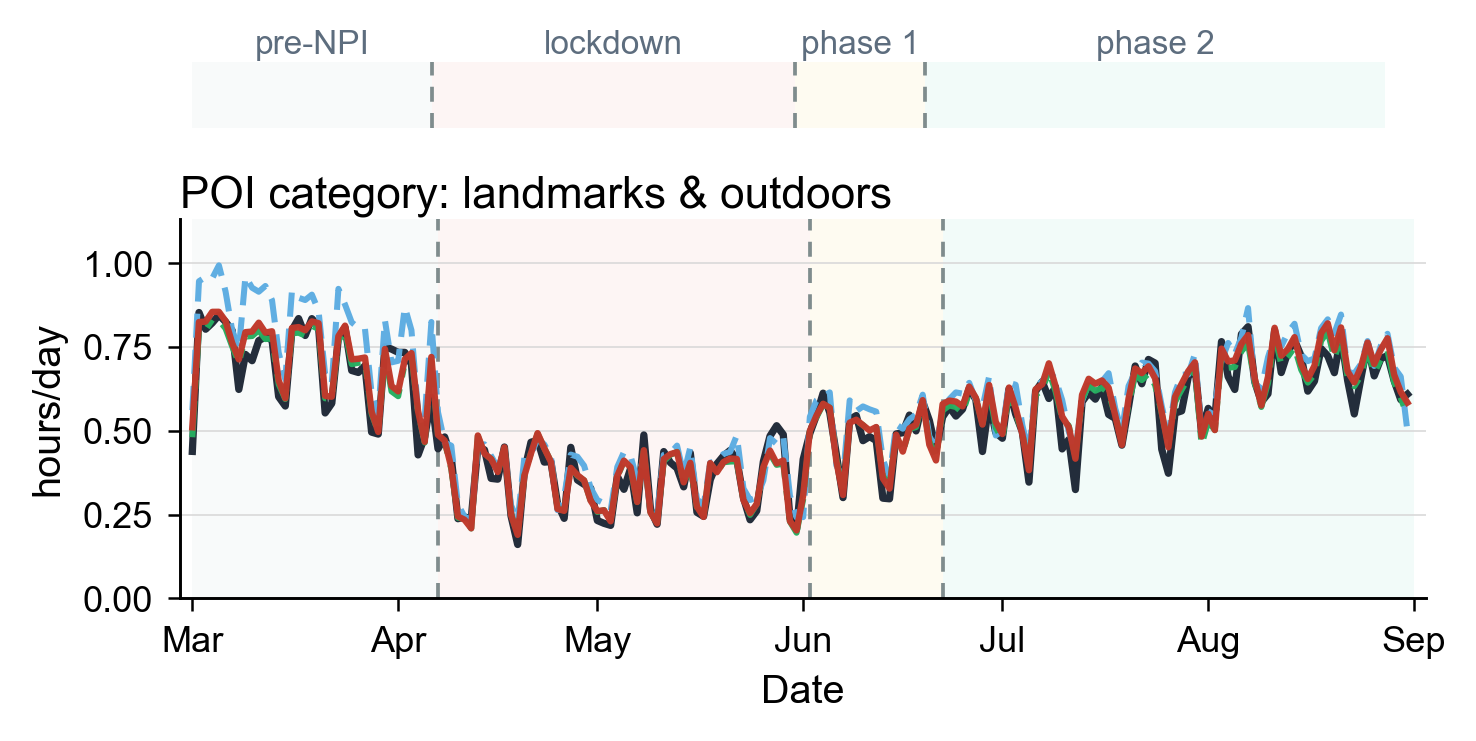}
\hfill
\includegraphics[width=0.48\textwidth]{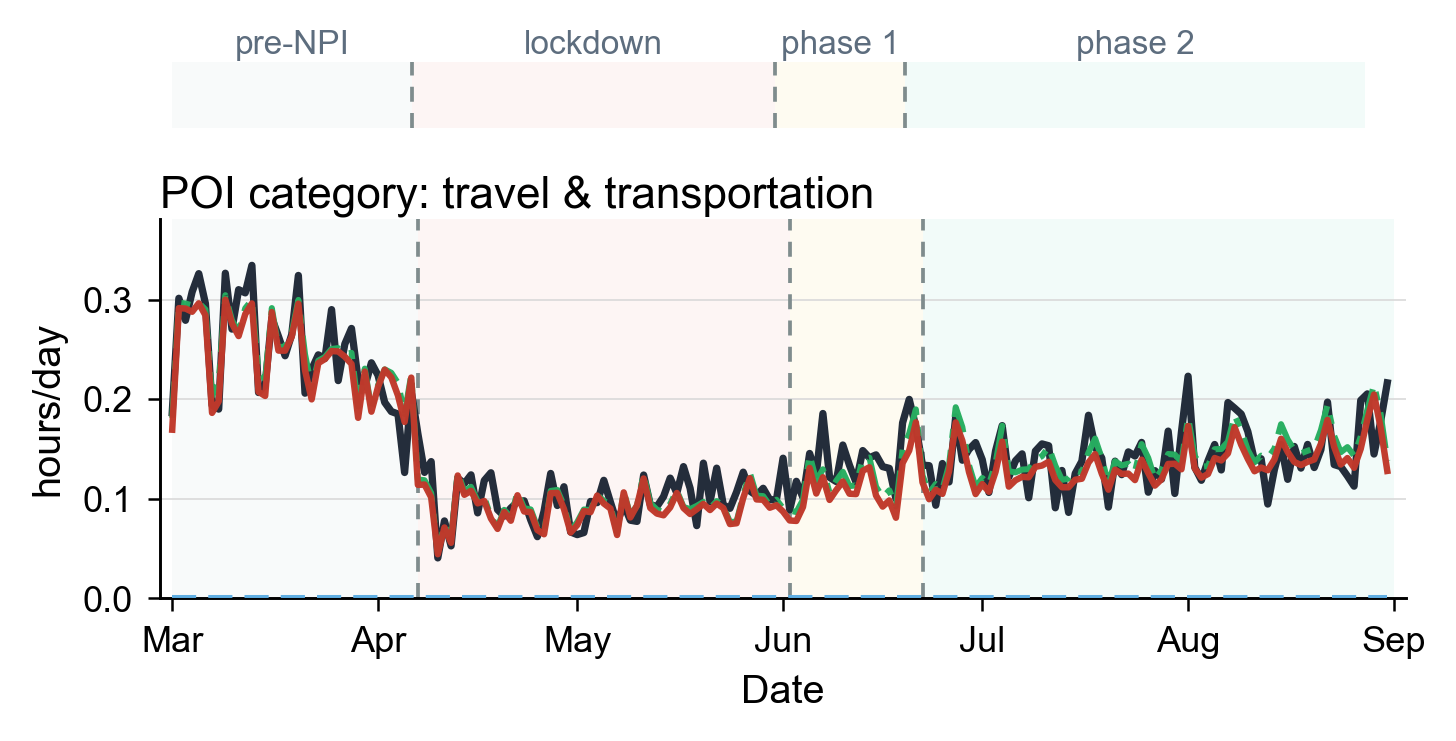}

\medskip

\includegraphics[width=0.48\textwidth]{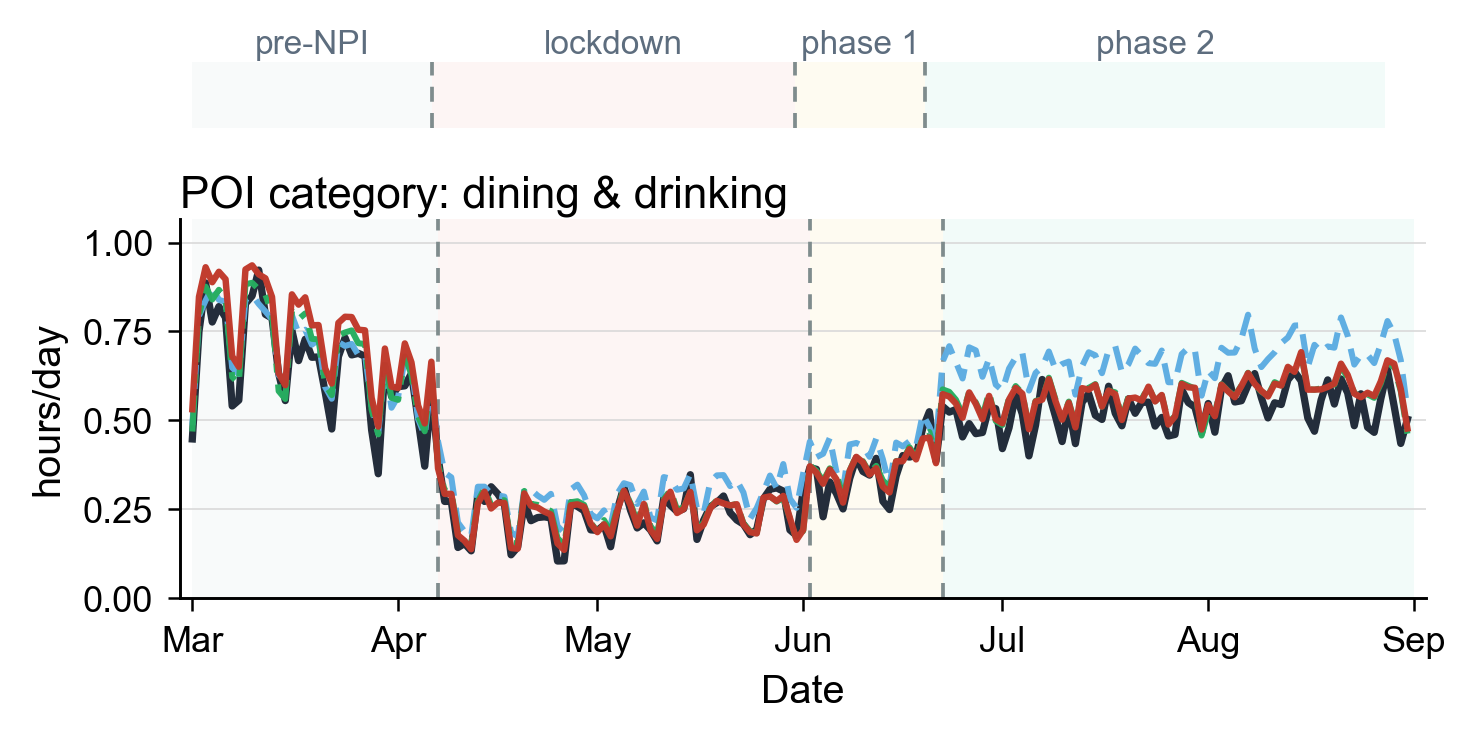}
\hfill
\includegraphics[width=0.48\textwidth]{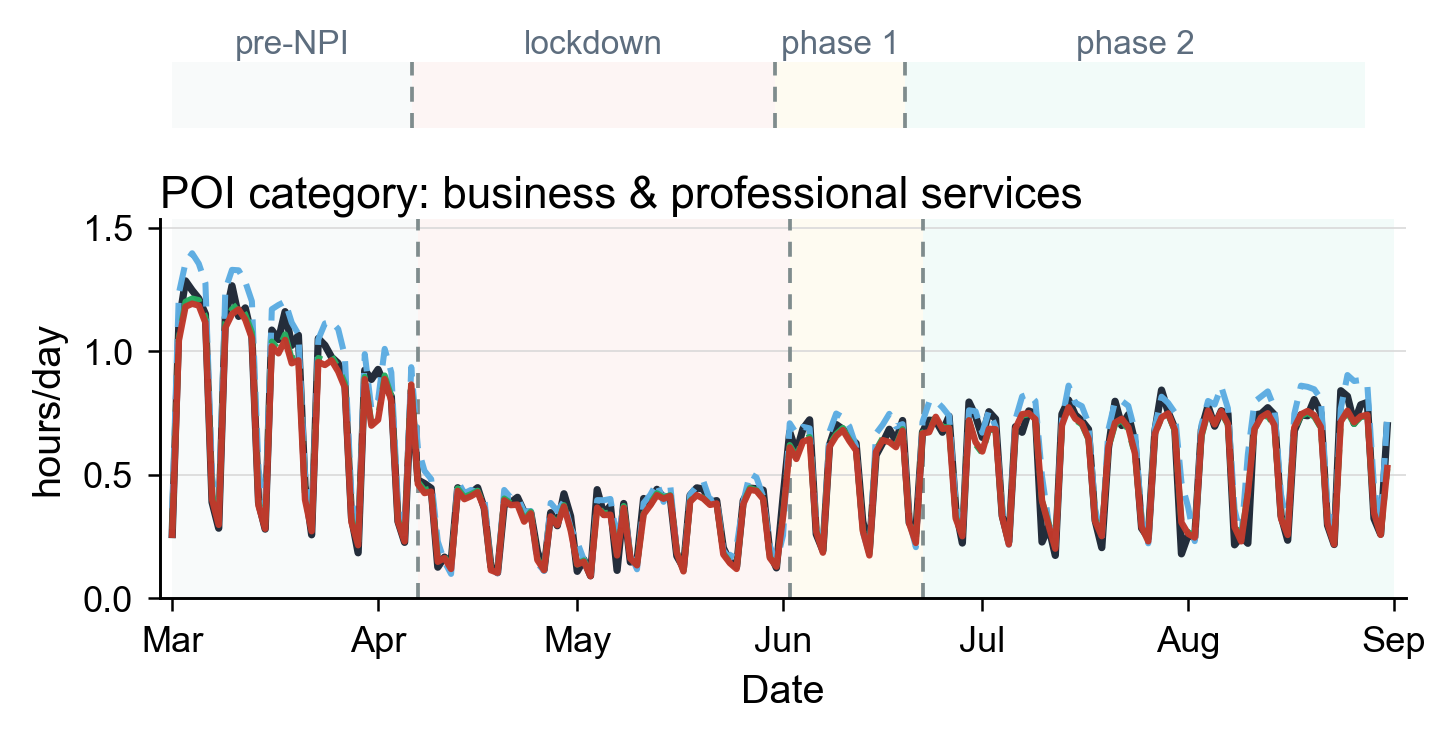}
\caption{Factual POI trajectories (Part~1).}
\label{fig:factual_daily_poi_profiles_a}
\end{figure*}

\begin{figure*}[t]
\centering
\includegraphics[width=0.78\linewidth]{figs/population_daily_factual_legend.png}

\medskip

\includegraphics[width=0.48\textwidth]{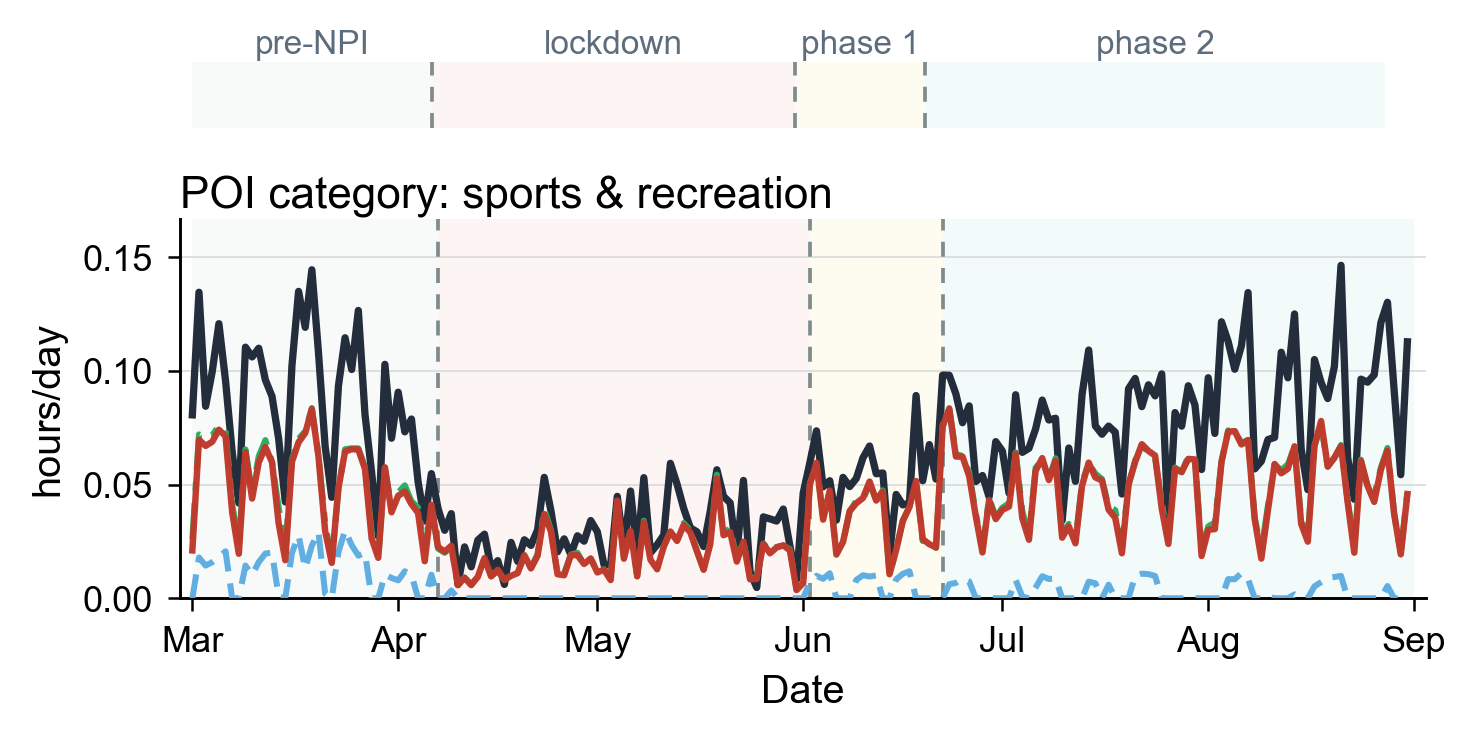}
\hfill
\includegraphics[width=0.48\textwidth]{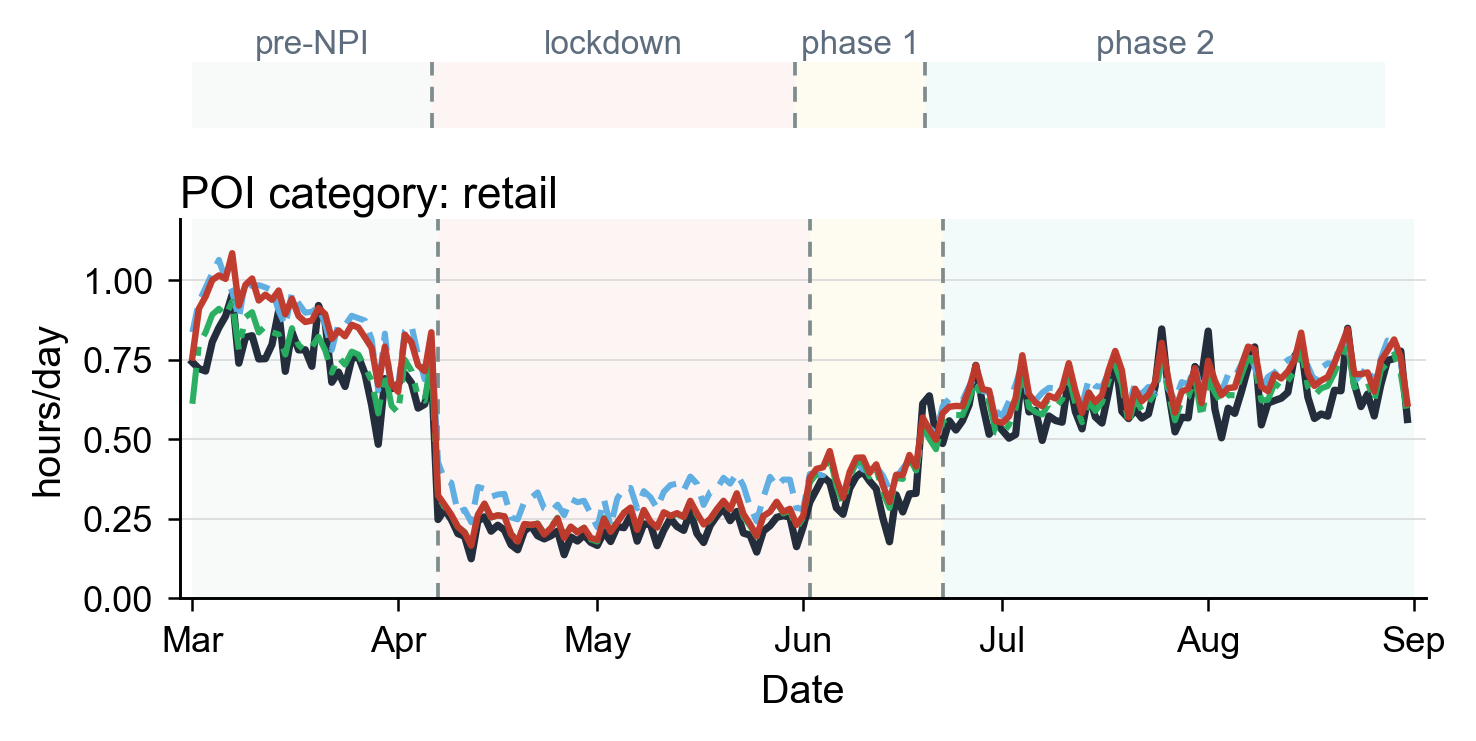}

\medskip

\includegraphics[width=0.48\textwidth]{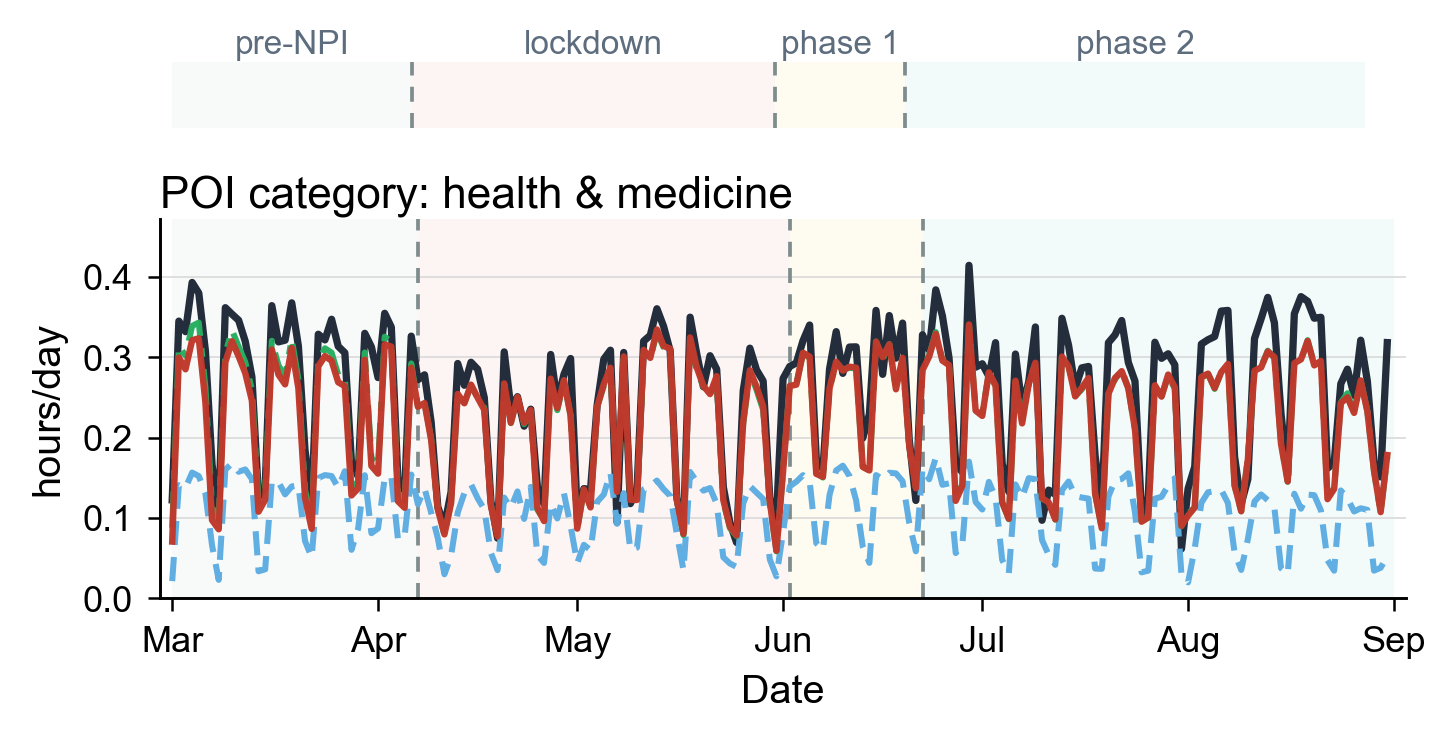}
\hfill
\includegraphics[width=0.48\textwidth]{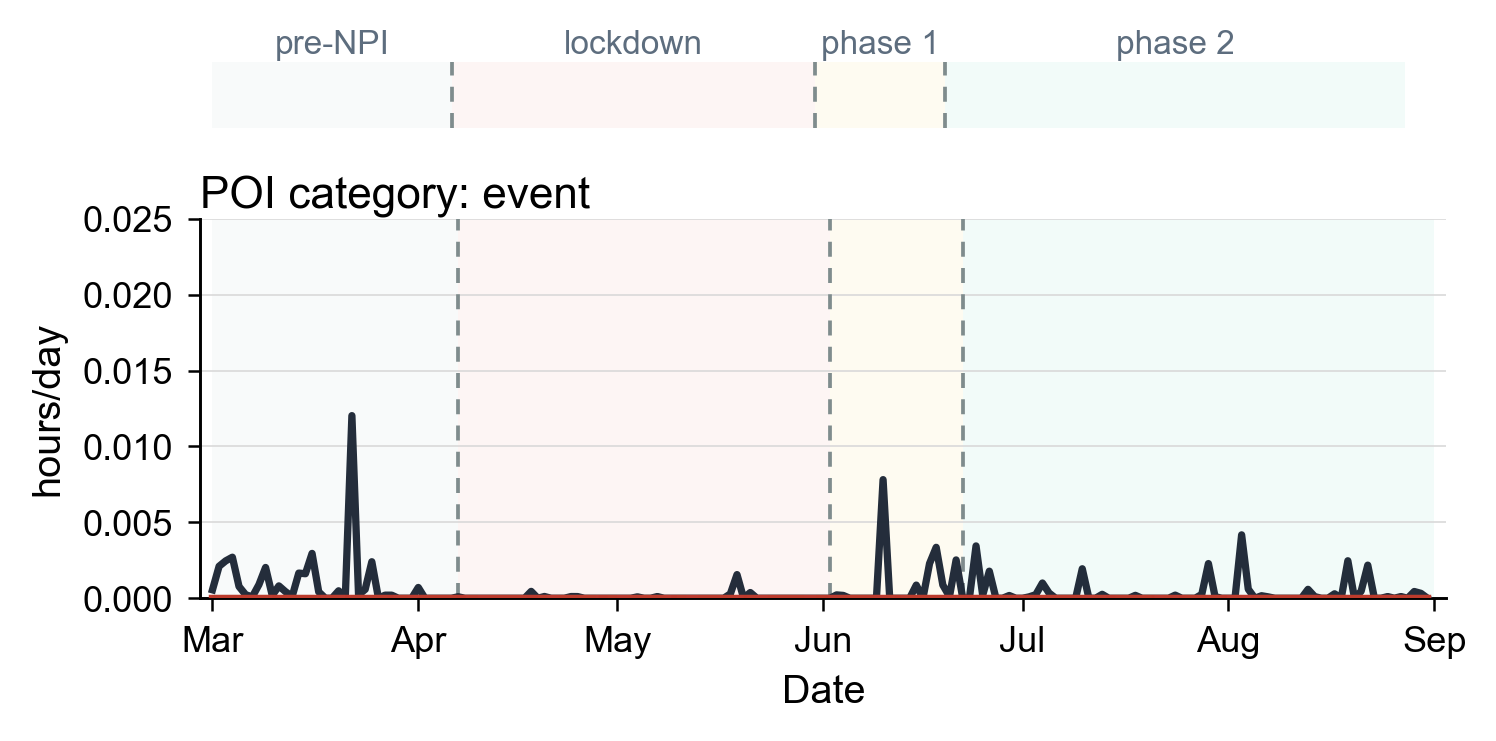}

\medskip

\includegraphics[width=0.48\textwidth]{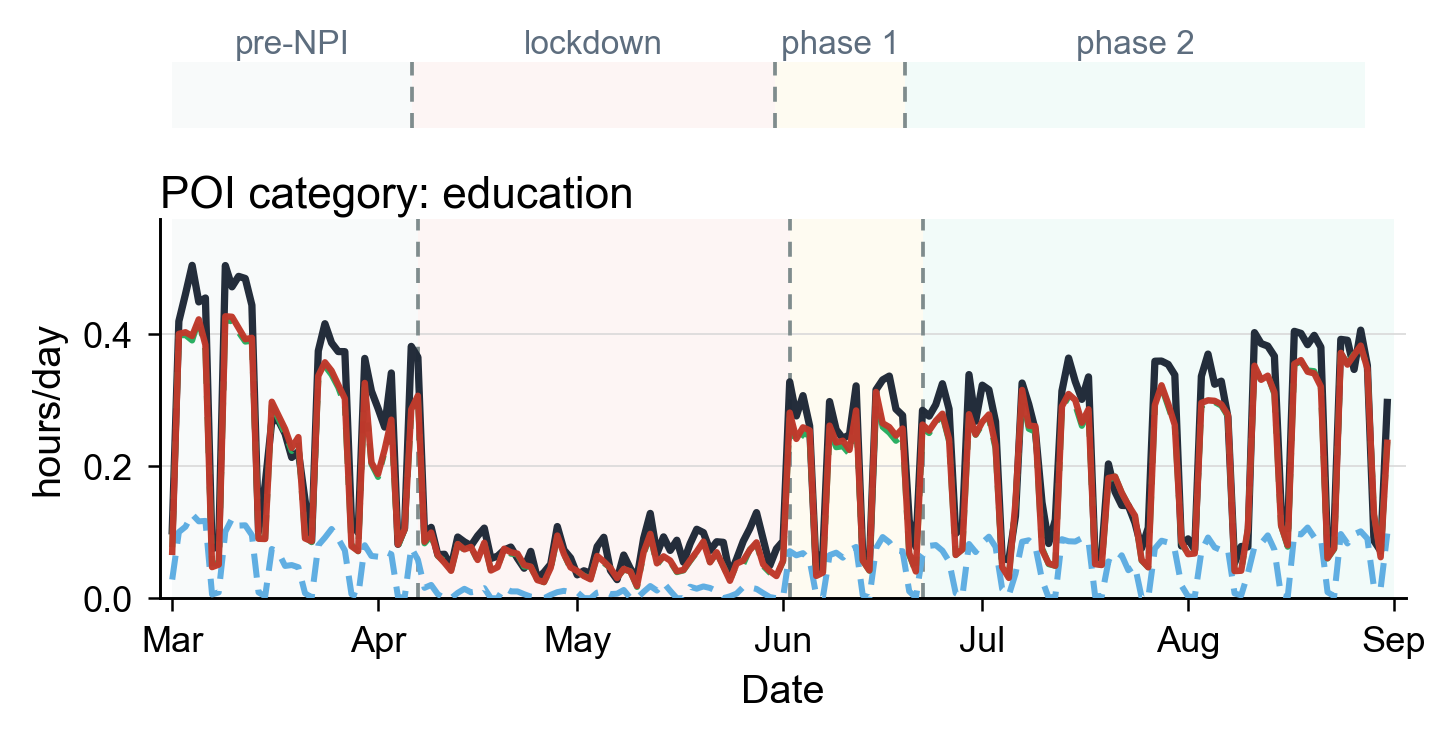}
\hfill
\includegraphics[width=0.48\textwidth]{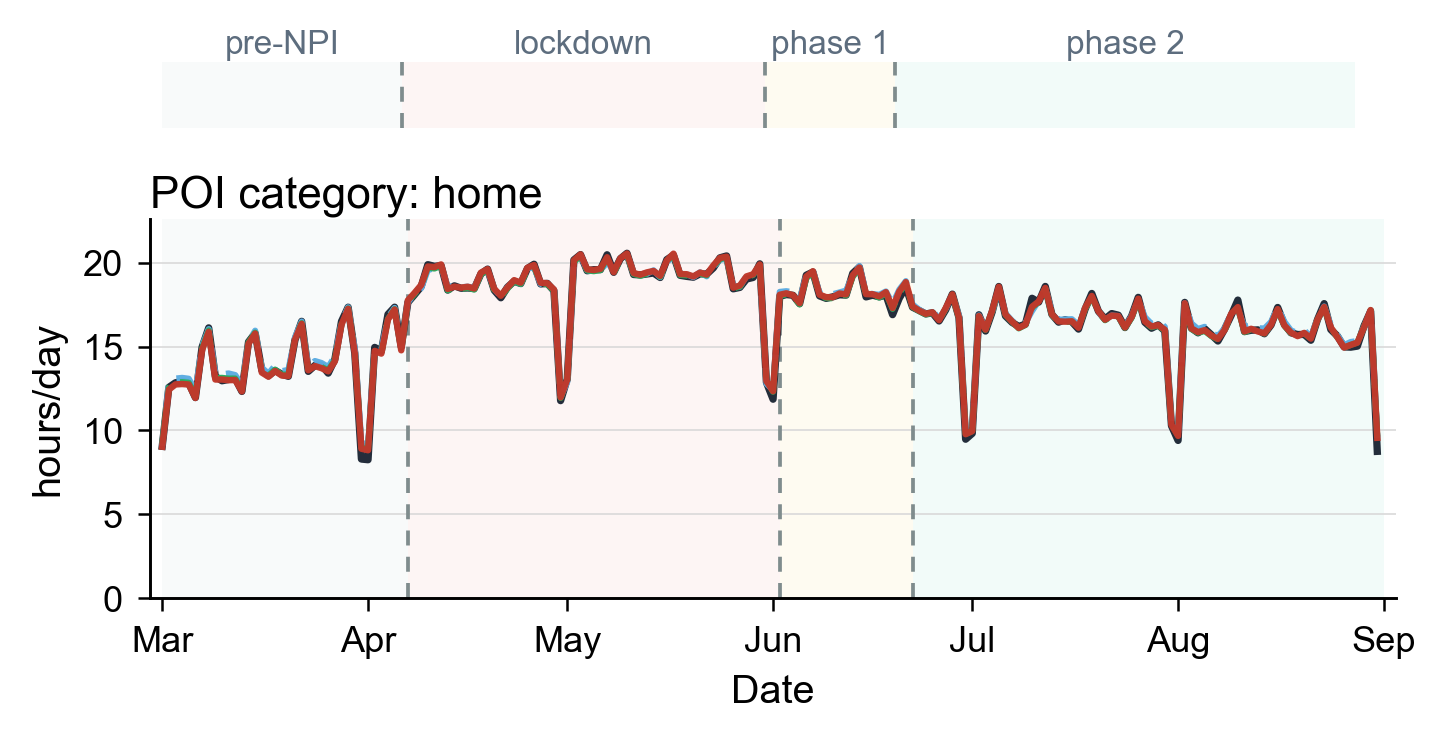}
\caption{Factual POI trajectories (Part~2).}
\label{fig:factual_daily_poi_profiles_b}
\end{figure*}

The daily trajectory figures make the behavioral mechanism visible in a way
that aggregate error metrics alone cannot. In
Figures~\ref{fig:factual_daily_poi_profiles_a} and
\ref{fig:factual_daily_poi_profiles_b}, the observed sequence exhibits a sharp
rise in home time and simultaneous reductions in discretionary out-of-home
categories during lockdown, followed by heterogeneous reopening patterns across
retail, dining and drinking, business and professional services, health, and
education. UrbanShare captures the timing of the major disruption but often
misstates the amplitude of policy-sensitive categories. UrbanShare-MoE reduces
these category-level distortions, and UrbanShare-MoE-PA preserves the same
improvement while producing smoother phase transitions. The rare categories,
especially event and arts/entertainment, remain more volatile because the
underlying observed hours are close to zero; we therefore interpret them mainly
as low-volume consistency checks rather than as central behavioral outcomes.

\begin{figure*}[t]
\centering
\includegraphics[width=0.78\linewidth]{figs/population_daily_factual_legend.png}

\medskip

\includegraphics[width=0.48\textwidth]{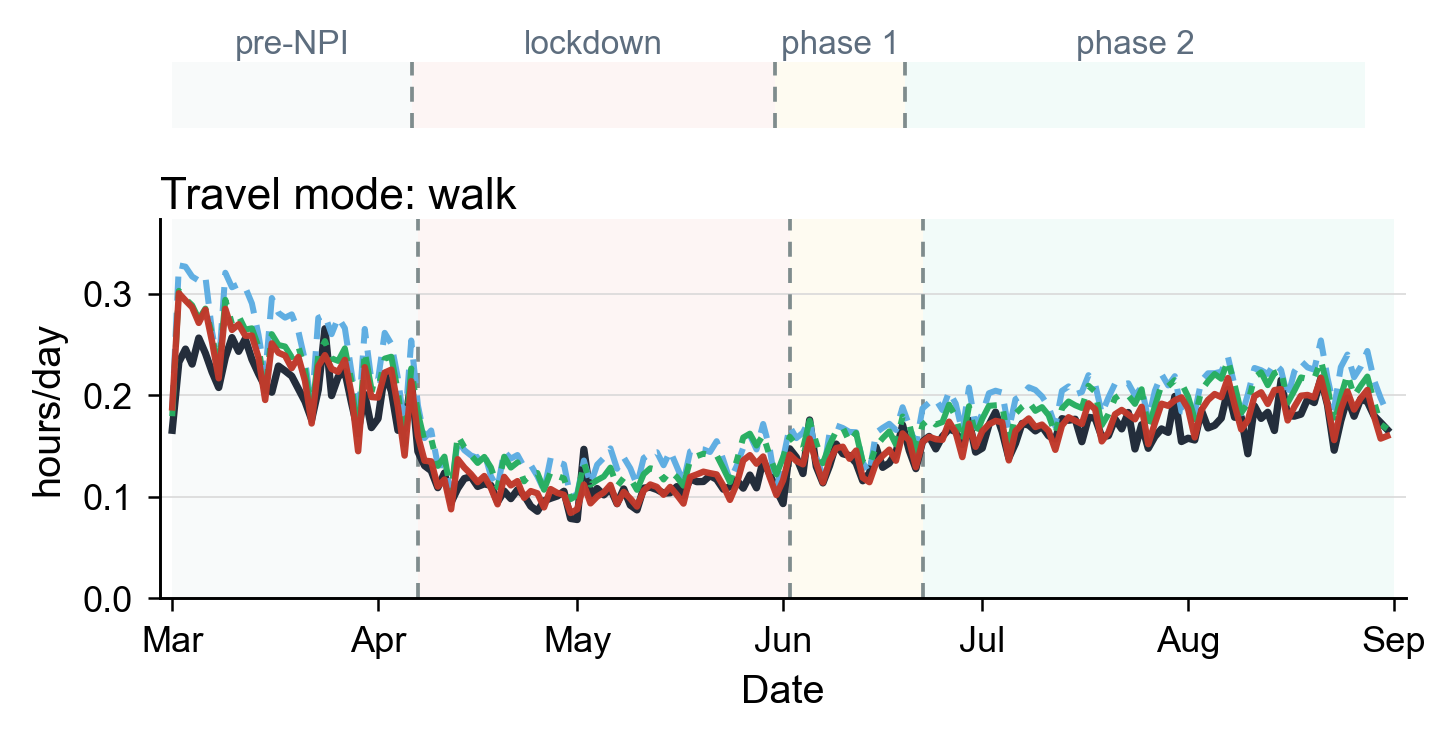}
\hfill
\includegraphics[width=0.48\textwidth]{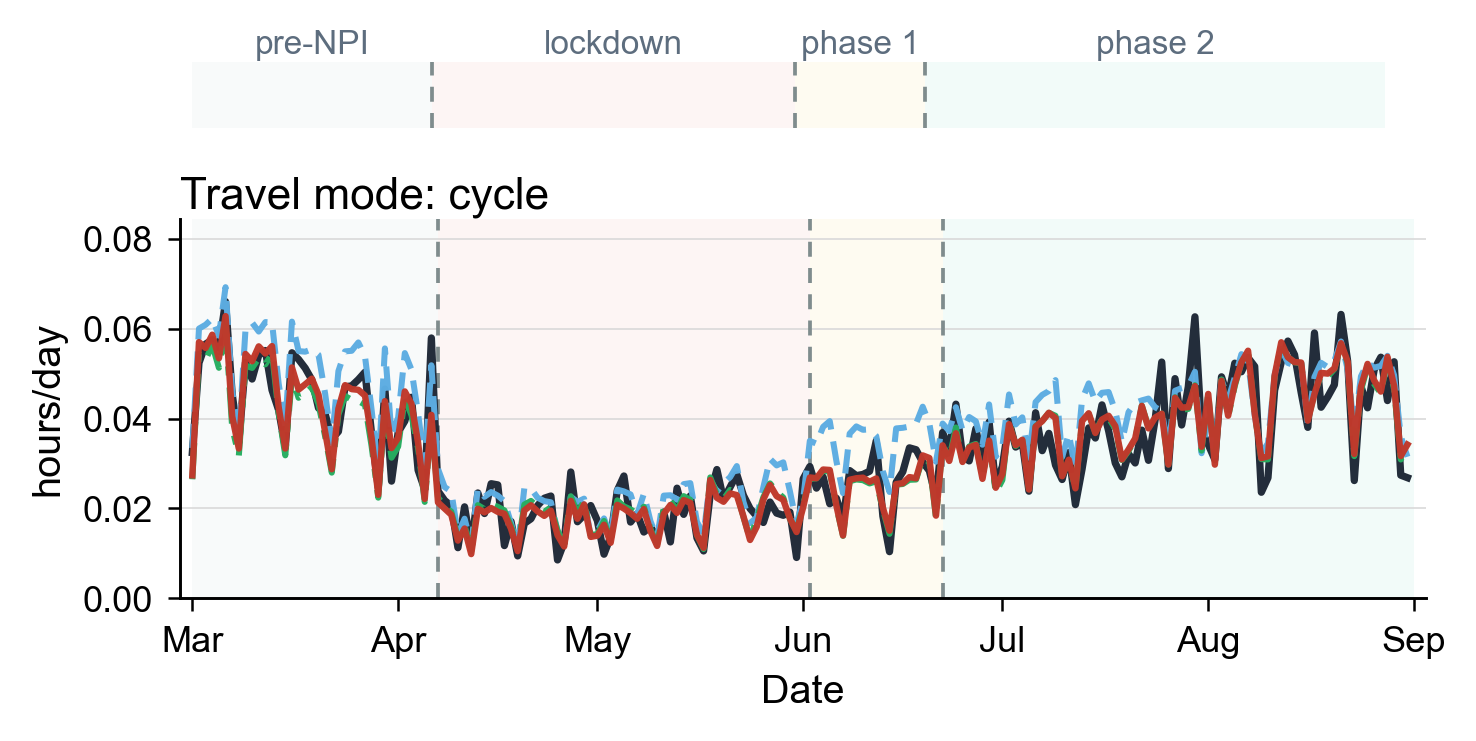}

\medskip

\includegraphics[width=0.48\textwidth]{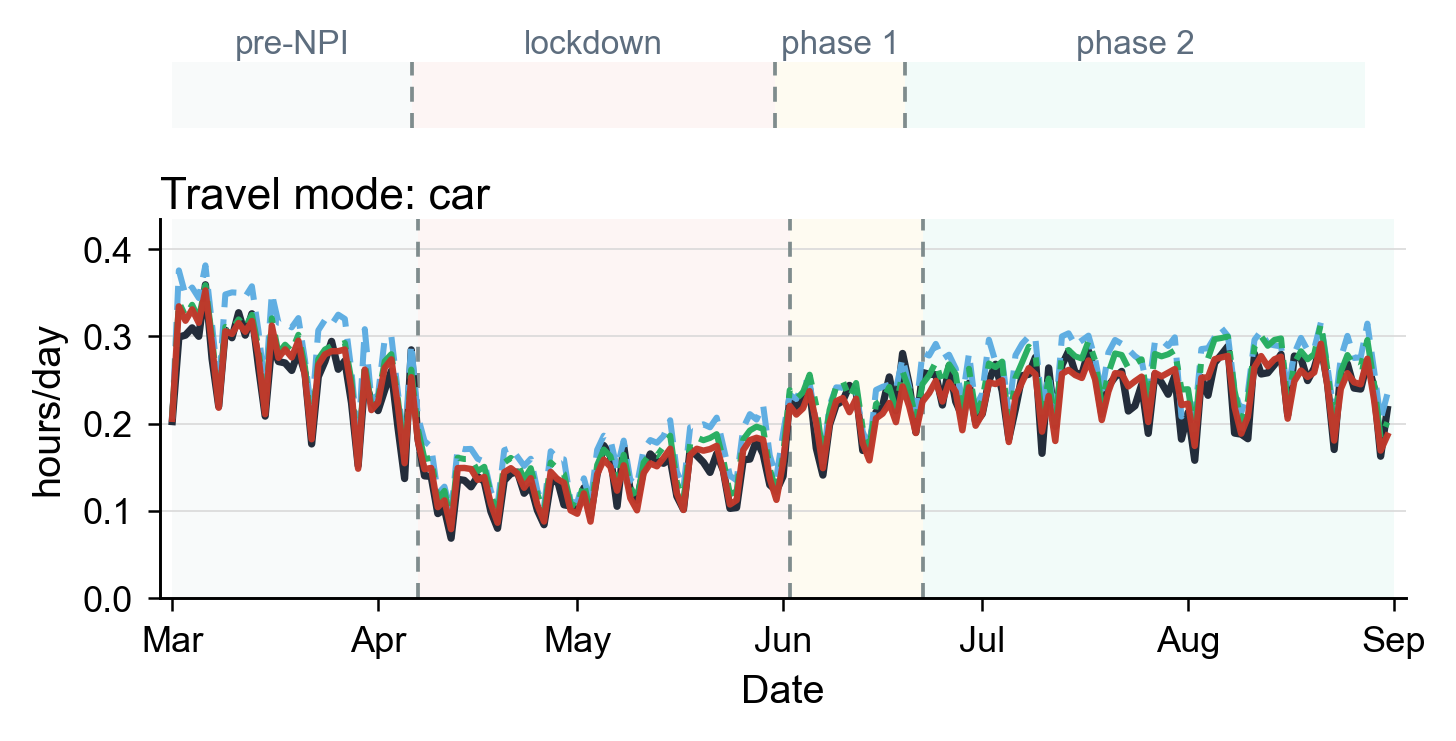}
\hfill
\includegraphics[width=0.48\textwidth]{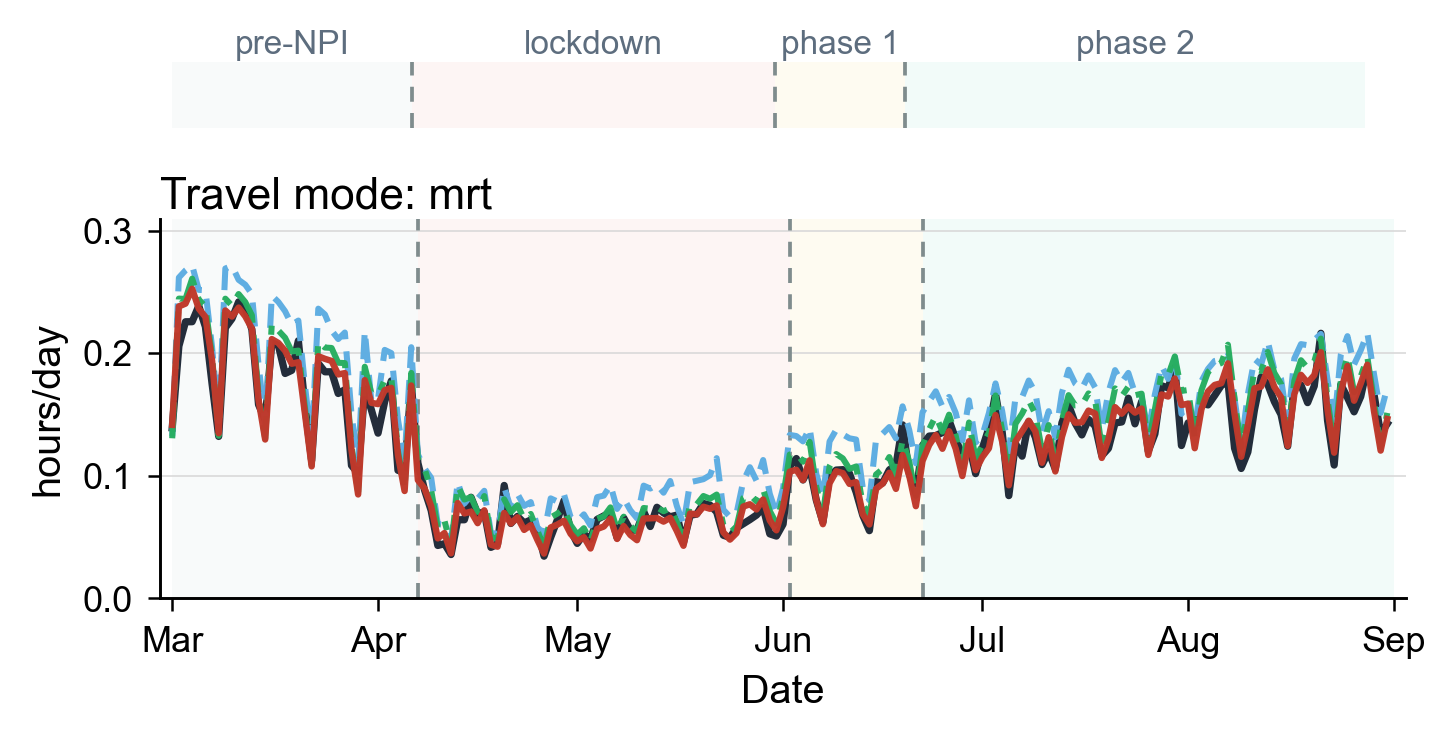}

\medskip

\makebox[\textwidth][c]{\includegraphics[width=0.48\textwidth]{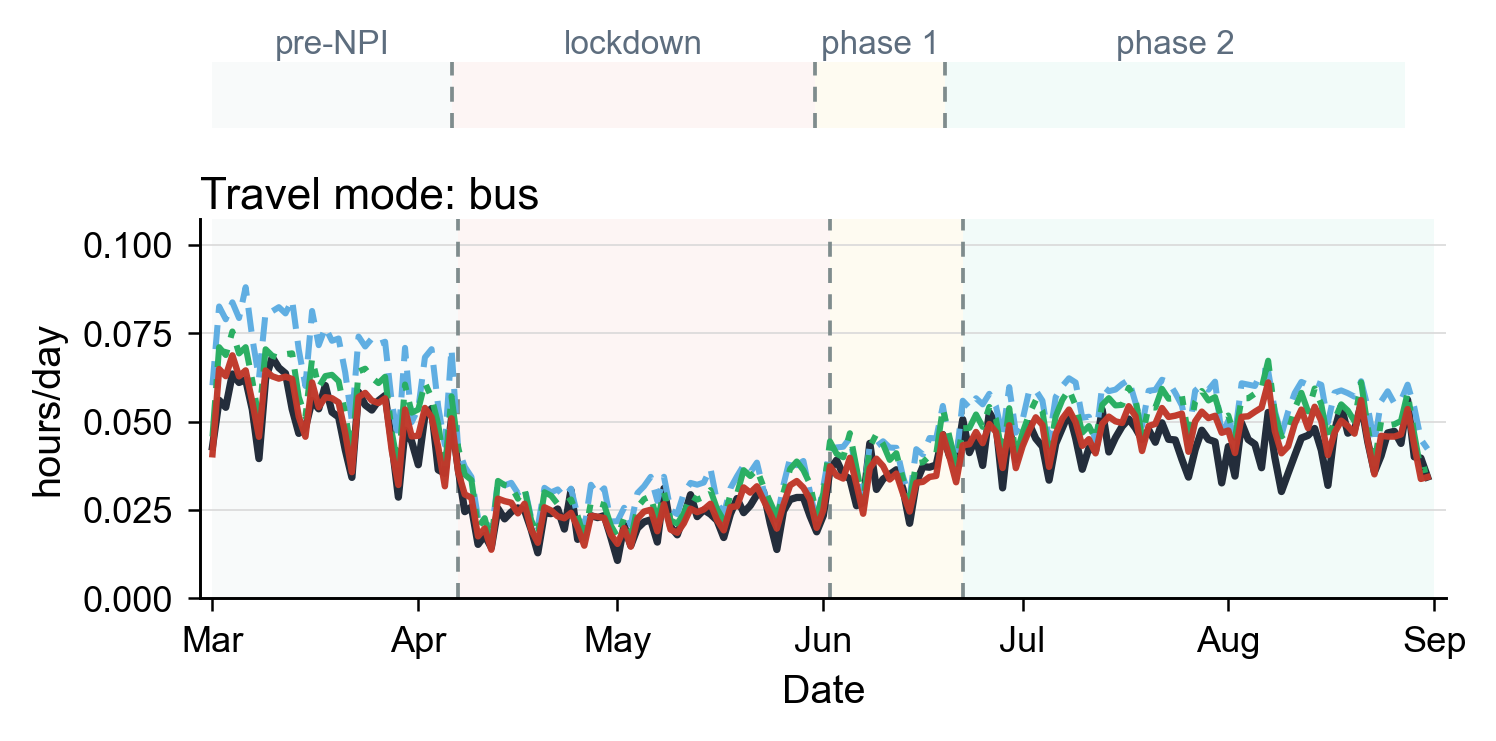}}
\caption{Factual travel-mode trajectories.}
\label{fig:factual_daily_mode_profiles}
\end{figure*}

\begin{figure*}[t]
\centering
\includegraphics[width=0.92\linewidth]{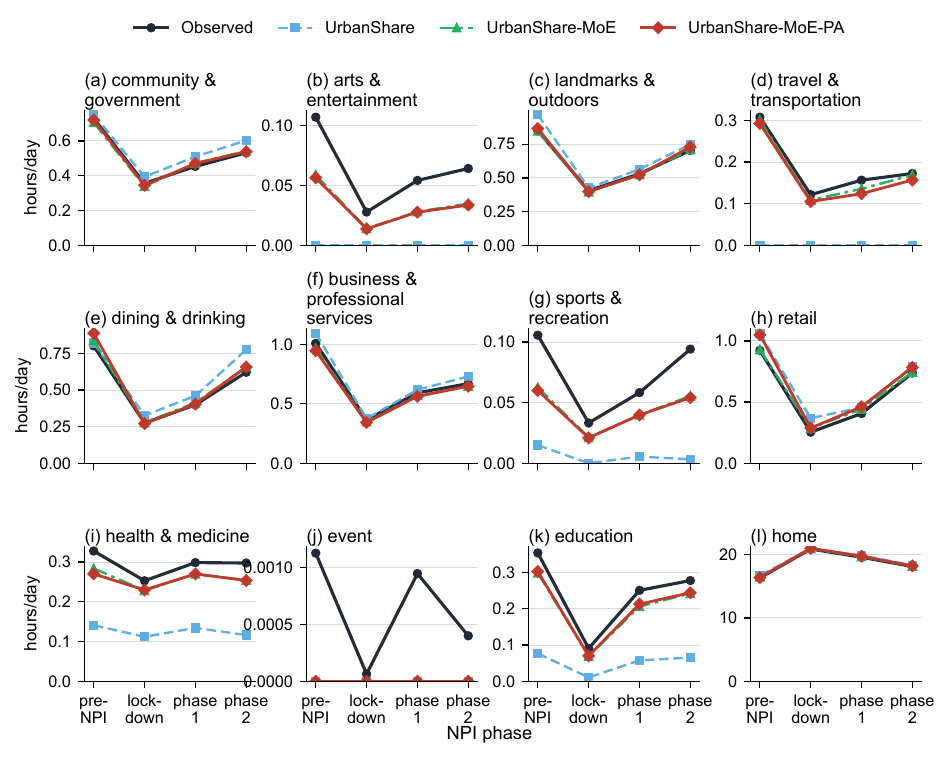}
\caption{Factual POI phase profiles.}
\label{fig:factual_phase_profiles_poi}
\end{figure*}

\begin{figure*}[t]
\centering
\includegraphics[width=0.72\linewidth]{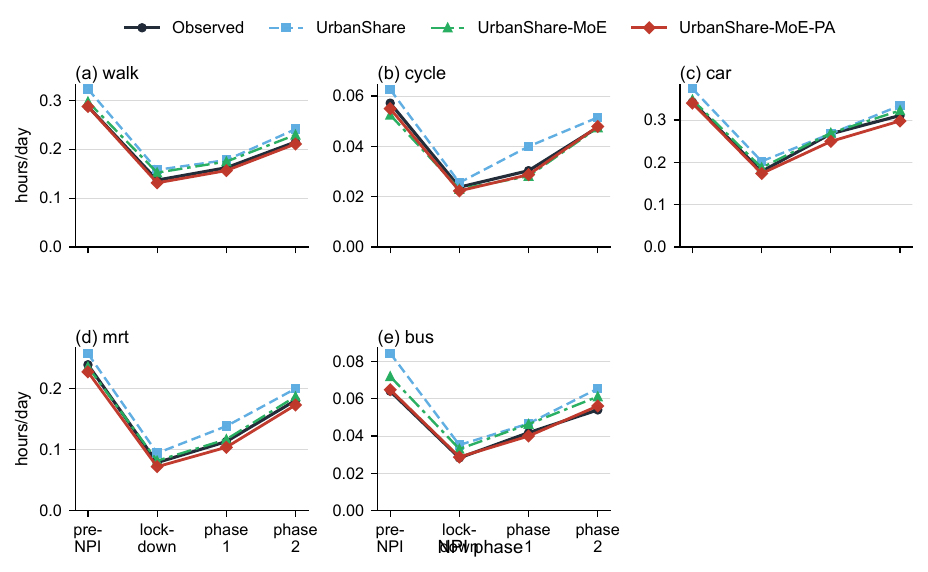}
\caption{Factual travel-mode phase profiles.}
\label{fig:factual_phase_profiles_mode}
\end{figure*}

Figure~\ref{fig:factual_daily_mode_profiles} shows the corresponding travel-mode
response. Walking, MRT, car, and bus all contract around lockdown and recover
after reopening, but the recovery is not uniform across modes. UrbanShare tends
to retain visible level bias in several modes, whereas UrbanShare-MoE-PA is
closest to the observed daily profiles, particularly for the public-transport
and walking series that are most sensitive to NPI timing. The phase summaries
in Figures~\ref{fig:factual_phase_profiles_poi} and
\ref{fig:factual_phase_profiles_mode} provide a compact view of the same
pattern: MoE routing improves POI-category reconstruction, while
preference-aligned training most clearly improves travel-mode calibration.

\subsection{MoE-PA Expert Profiles and Behavioral Heterogeneity}
\label{sec:results_experts}
\begin{figure*}[t]
\centering
\includegraphics[width=0.78\linewidth]{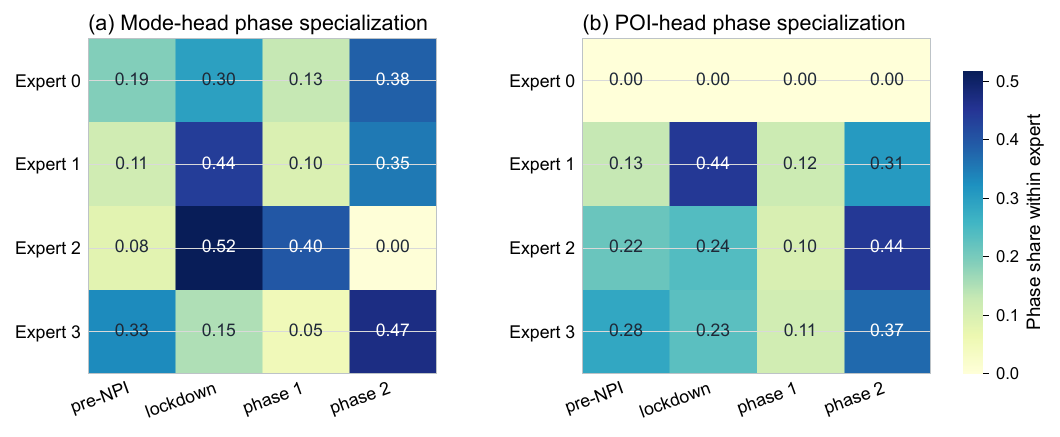}
\caption{Expert phase specialization.}
\label{fig:moe_phase_specialization}
\end{figure*}

To understand how the proposed UrbanShare-MoE-PA model represents behavioral
heterogeneity, we inspect the routed experts under the factual
UrbanShare-MoE-PA checkpoint itself rather than under the earlier
supervised-only MoE model. As a compact overview, we retain the
phase-specialization view in Figure~\ref{fig:moe_phase_specialization}, which
shows how routed experts are distributed across the four NPI phases. The phase
pattern is already structured: the mode head contains two reopening-oriented
experts and two more lockdown-oriented experts, while the POI head separates a
nearly inactive expert from a strongly home-centered lockdown expert and two
more open phase-2-oriented experts.

\begin{figure*}[t]
\centering
\begin{subfigure}[t]{0.38\linewidth}
\centering
\includegraphics[width=\linewidth]{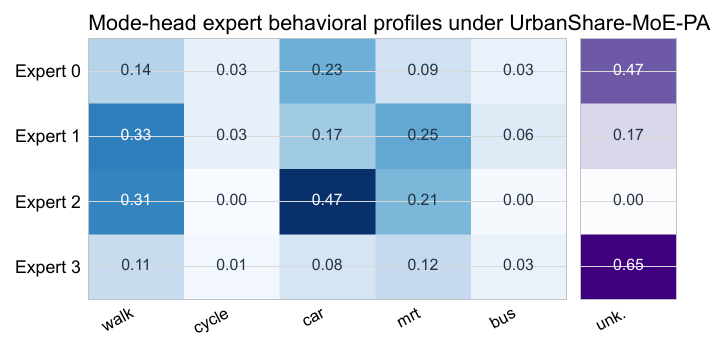}
\caption{Mode experts.}
\end{subfigure}
\hfill
\begin{subfigure}[t]{0.58\linewidth}
\centering
\includegraphics[width=\linewidth]{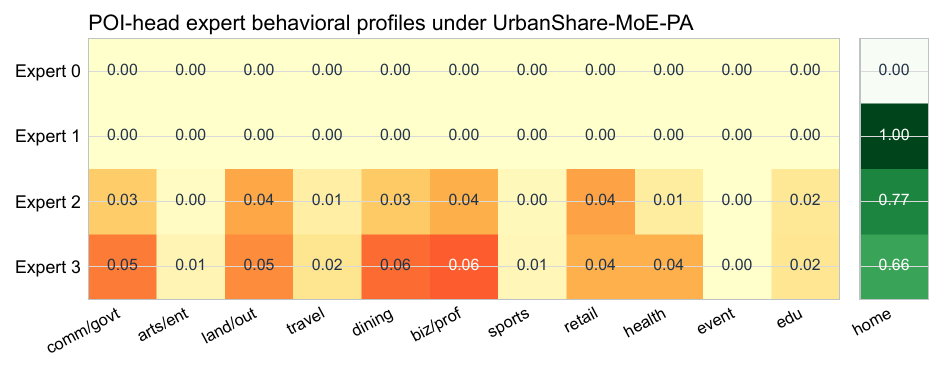}
\caption{POI experts.}
\end{subfigure}
\caption{Expert behavioral profiles.}
\label{fig:moe_profiles}
\end{figure*}

Figures~\ref{fig:moe_profiles}(a)--(b) then visualize the corresponding
behavioral profiles. On the mode side, Expert~0 is the largest reopening
expert (45.7\% of routed samples) and captures substantial everyday mobility,
with a mixed composition dominated by unknown mode, car, and walk. Expert~3 is
also reopening-oriented but exhibits a distinct lower-mobility
unknown-plus-MRT structure. By contrast, Experts~1 and~2 are much more
lockdown-oriented and correspond to near-zero-mobility samples, so their mode
compositions should be interpreted as sparse residual travel rather than as
full mobility archetypes. This indicates that the mode head separates
restricted-mobility and reopening-mobility regimes, while still preserving
finer distinctions within the reopening stage.

On the POI side, the heterogeneity pattern is even clearer. Expert~1 is a
home-centered lockdown expert, receiving 32.3\% of routed samples with an
almost pure home allocation. Expert~2 is the dominant reopening expert
(48.4\%), combining a lower home share with larger landmarks/outdoors,
business/professional services, retail, and dining allocations. Expert~3 is a
second reopening profile with relatively stronger community/government, dining,
business/professional, and health-related activity, while Expert~0 is
effectively inactive. Taken together, these results show that the PA-tuned MoE
does not merely improve aggregate fidelity: it routes different agent-day observations to distinct travel-mode and
POI-category regimes. This expert-level structure is consistent with
heterogeneous behavioral regimes across agent-days and helps explain the stronger factual
reconstruction results reported in Section~\ref{sec:results_factual}.

\subsection{Alternative-Calendar Behavioral Trajectories}
\label{sec:results_calendar}
\begin{figure*}[t]
\centering
\includegraphics[width=0.92\linewidth]{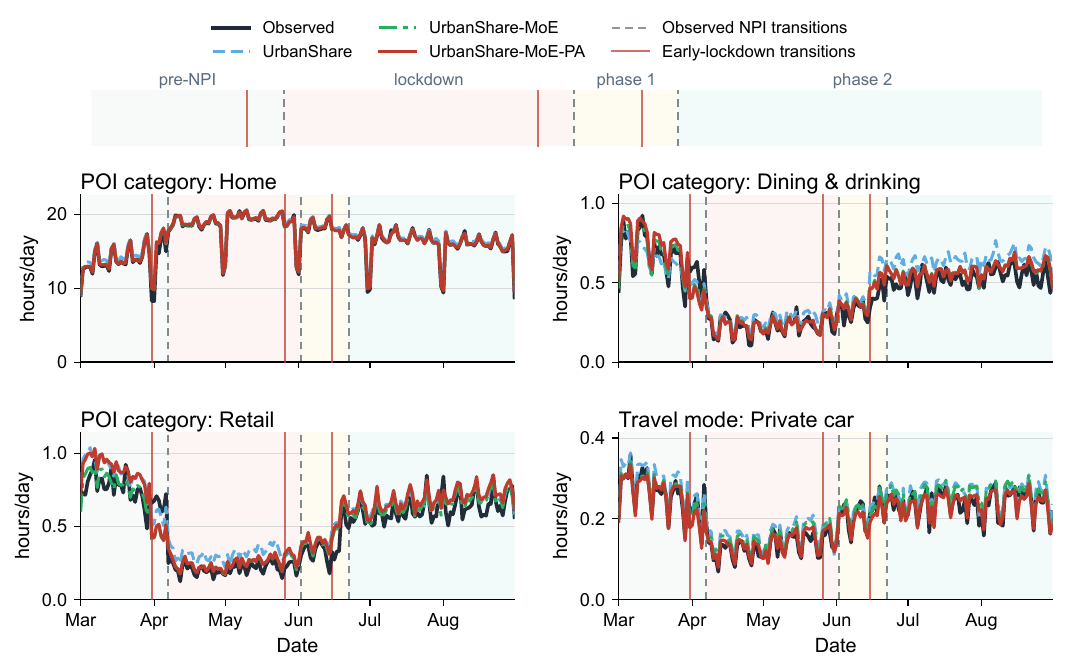}
\caption{Early-lockdown behavioral trajectories.}
\label{fig:early_daily_cf_profiles}
\end{figure*}

Because the alternative calendars are not historically observed, this part of the
analysis focuses on behavioral trajectory comparison. To keep the main text
concise, we use the early-lockdown scenario as the representative behavioral
example here,
while Section~\ref{sec:results_seir} presents calibrated epidemic comparisons and
Section~\ref{sec:results_tradeoff} reports epidemic--activity trade-offs across all four calendars. To make the
calendar shift readable, Figure~\ref{fig:early_daily_cf_profiles} focuses on
daily population-mean trajectories for representative POI and travel-mode
series rather than on dense calendar-wide panels. The black curve provides the
observed factual reference, while the colored curves show the early-lockdown
outputs of UrbanShare, UrbanShare-MoE, and UrbanShare-MoE-PA under the shifted
calendar.

Under the simulation design in Section~\ref{sec:cf}, these early-lockdown panels
represent full calendar-conditioned re-simulations rather than a post-31-March
splice. Because the shifted calendar changes both the phase labels and the
phase-relative timing features that condition the behavior model, small
pre-boundary deviations can appear on the absolute date axis.

The qualitative response is directionally coherent with the shifted policy
calendar. Relative to the observed sequence, the alternative-calendar runs shift forward
the main home-centered transition by approximately one week. Dining and drinking, as well as retail activity, contract earlier, remain lower
during the shifted lockdown interval,
and begin to recover earlier once the alternative-calendar reopening begins. The car
series shows a milder contraction, because private vehicle use is less directly constrained than public transport
use and many discretionary out-of-home activities. These daily curves therefore show not only that activity levels
change, but also that the timing of behavioral adjustment moves with the
alternative calendar.

The three behavior models, however, do not respond equally well to the shifted
calendar. UrbanShare captures the broad direction of change but still exhibits
noticeable level distortions in both POI and mode intensities. UrbanShare-MoE
produces a cleaner contraction-and-rebound structure, indicating that expert
routing helps preserve distinct phase-sensitive response patterns under the alternative-calendar
calendar. UrbanShare-MoE-PA yields the most coherent temporal shift: the main
behavioral break aligns most clearly with the shifted policy boundaries, while
the subsequent reopening recovery is advanced without introducing abrupt discontinuities. UrbanShare-MoE-PA is therefore carried forward as the primary trajectory for the
subsequent epidemic and activity comparison.

\subsection{Trend-Calibrated SEIR Results}
\label{sec:results_seir}
Before using the epidemic simulator for scenario comparison, we first calibrate
the SEIR parameters under the observed factual behavioral profile. Because the
simulator operates on 911 sampled agents rather than the full Singapore
population, the goal is not to reproduce raw national counts. Instead, the
calibration targets scale-free epidemic shape descriptors. Specifically, we
compare normalized cumulative infections $(N-S_t)$ against the official
national cumulative case curve and compare simulated new exposures against the
official 7-day-smoothed reported daily case curve after a fixed 21-day lead
alignment \citep{moh2020pastupdates,cda2025covid}. This alignment reflects the
gap between latent infection and reported case observation and is used only for
trend calibration, not for downstream scenario evaluation.

Table~\ref{tab:seir_trend_calibration} reports the diagnostics for the
automatically calibrated factual SEIR run. The calibrated simulator attains a
lead-aligned incidence correlation of 0.59 with normalized RMSE 0.31, while the
cumulative-curve comparison reaches a correlation of 0.70 with normalized RMSE
0.42. The first-wave peak-date error is 1 day and the phase-share MAE is 0.19.
These diagnostics show that the calibrated SEIR captures the timing and broad
first-wave shape of Singapore's epidemic at the trend level, while deliberately
avoiding a literal fit to national case counts.

\begin{figure*}[t]
\centering
\includegraphics[width=0.98\linewidth]{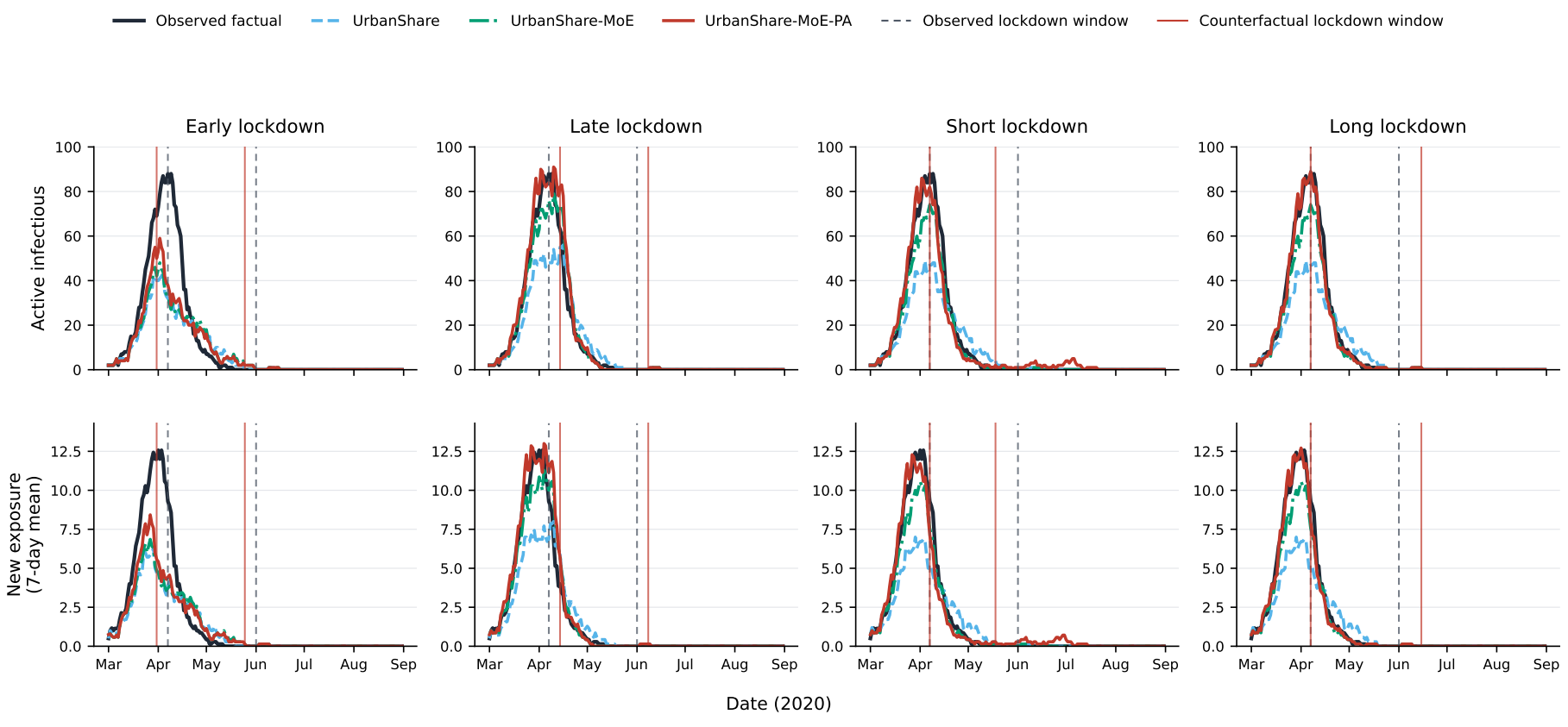}
\caption{SEIR policy comparisons.}
\label{fig:seir_policy_compare}
\end{figure*}

Figure~\ref{fig:seir_policy_compare} propagates the factual trajectory and the
alternative-calendar trajectories from all three behavioral models through the
same calibrated SEIR parameters. The main SEIR configuration uses the fixed
background infection probability, population-mean home protection, and
crowding elasticity $\xi=1$ described in Section~\ref{sec:seir}. Across all
four calendars, UrbanShare produces strongly attenuated epidemic curves because
its alternative-calendar mobility trajectories over-contract non-home activity.
UrbanShare-MoE restores part of the first-wave shape but produces less separation
across policy calendars. UrbanShare-MoE-PA yields the clearest temporal
differentiation across calendars and retains the timing structure induced by the
corresponding behavioral rollouts. It is therefore used as the primary trajectory
for the policy trade-off analysis.

The four calendar panels show simulated responses that are consistent with the
intended intervention logic. Under early lockdown, the UrbanShare-MoE-PA curve
turns downward earlier than the factual trajectory and the main infectious peak
is substantially reduced. In the calibrated simulator, this pattern corresponds
to lower exposure accumulation before the first-wave peak. Under late lockdown,
the curve remains elevated for longer and reaches a slightly higher peak than
the factual trajectory, consistent with a delayed mobility contraction in the
generated behavior. Under short lockdown, the first peak remains close to the
factual trajectory, but the earlier release leaves a small post-lockdown tail.
Under long lockdown, the peak is not substantially reduced because the observed
peak occurs near the beginning of the lockdown window; instead, the longer
restriction period mainly suppresses the post-peak tail and lowers cumulative
burden. This pattern supports the internal logic of the simulator: earlier
timing affects the first-wave peak, whereas longer duration mainly affects the
post-peak tail.

Some separation between factual and alternative-calendar epidemic curves appears
before the historical lockdown boundary on the absolute date axis. This pattern
is inherited from the calendar-conditioned behavioral rollout in
Section~\ref{sec:cf}, where shifted phase-relative features and recursively
updated behavioral history propagate into the exposure states.

\begin{table}[t]
\centering
\caption{SEIR trend-calibration diagnostics.}
\label{tab:seir_trend_calibration}
\begin{tabular}{lc}
\toprule
Metric & Value \\
\midrule
Lag-adjusted incidence correlation & 0.59 \\
Lag-adjusted incidence NRMSE & 0.31 \\
First-wave peak-date error (days) & 1 \\
Phase-share MAE & 0.19 \\
Cumulative-curve correlation & 0.70 \\
Cumulative-curve NRMSE & 0.42 \\
\bottomrule
\end{tabular}
\end{table}

\subsection{Cross-Calendar Epidemic and Activity Trade-Offs}
\label{sec:results_tradeoff}
Table~\ref{tab:policy_tradeoff_models} summarizes the downstream epidemic and
activity outcomes for UrbanShare-MoE-PA, using the original policy sequence as
the within-model reference. This specification combines low factual
reconstruction errors with the clearest differentiation among the
alternative-calendar behavioral and SEIR trajectories.

The epidemic metrics show a clear timing pattern. The early-lockdown calendar
produces the largest simulated epidemiological improvement: relative to the
original policy, Peak $I$ falls from 83 to 59, Area($I$) falls from 2133 to
1615, and Final~$R$ falls from 305 to 231. This is the most favorable
disease-control scenario in the calibrated simulator because the intervention
starts before the first-wave peak. The late-lockdown calendar has the opposite
profile, increasing Peak $I$ to 91, Area($I$) to 2574, and Final~$R$ to 368.
This deterioration is consistent with the visual pattern in
Figure~\ref{fig:seir_policy_compare}, where delayed restrictions leave the
rising phase less constrained before lockdown takes effect in the simulated
trajectory.

Changing the duration while keeping the original start date produces a different
kind of trade-off. The short-lockdown calendar preserves more activity but
modestly increases epidemic burden, with Peak $I$ of 86 and Final~$R$ of 325.
The long-lockdown calendar reduces the post-peak burden relative to the
short-lockdown calendar, lowering Final~$R$ to 319 and Area($I$) to 2231, but it
remains less favorable than the original policy on the main epidemic summaries.
Thus, duration changes mainly reshape the post-peak tail, whereas start-date
changes reshape the first-wave peak itself.

The sector-weighted activity index shows the corresponding retained-activity pattern. Among the four
alternative calendars, short lockdown retains the most activity
($O=255.79$, $+1.06\%$ relative to the original policy), while late lockdown is
nearly unchanged ($O=253.11$, $-0.01\%$). Early and long lockdown have lower activity-index values
($O=247.00$ and $O=246.95$), reflecting stronger simulated reductions in
weighted urban activity during the intervention window. The resulting
scenario ordering is therefore not one-dimensional: early lockdown is strongest
under an epidemic-suppression objective; short lockdown is the most
activity-preserving alternative scenario but permits a modest increase in epidemic
burden; late lockdown preserves activity but performs poorly epidemiologically;
and long lockdown extends restrictions without outperforming the original policy
on either weighted activity or the main epidemic summaries. In this calibrated
setting, earlier timing yields a stronger simulated first-wave reduction than
simply extending the lockdown after the peak has already emerged.

\begin{table*}[t]
\centering
\caption{Epidemic and activity trade-offs.}
\label{tab:policy_tradeoff_models}
\scriptsize
\resizebox{\textwidth}{!}{%
\begin{tabular}{lrrrrrrr}
\toprule
Policy & Peak $I$ & Peak new $E$ & Final $R$ & Area($I$) & Area(new $E$) & $O$ & Change in $O$ (\%) \\
\midrule
Original policy & 83 & 11.86 & 305 & 2133 & 303 & 253.12 & -- \\
Early lockdown & 59 & 8.43 & 231 & 1615 & 229 & 247.00 & -2.42 \\
Late lockdown & 91 & 13.00 & 368 & 2574 & 366 & 253.11 & -0.01 \\
Short lockdown & 86 & 12.29 & 325 & 2273 & 323 & 255.79 & +1.06 \\
Long lockdown & 89 & 12.71 & 319 & 2231 & 317 & 246.95 & -2.44 \\
\bottomrule
\end{tabular}}
\end{table*}

\section{Discussion}
\label{sec:discussion}

\subsection{Mobility Response as Structured Daily Time Allocation}
The first implication of this study is that pandemic mobility response should be
interpreted as a reallocation of daily time, rather than as a scalar reduction in
movement. This finding is consistent with the activity-based view in transport
research, where travel is derived from activity participation and constrained by
time budgets, routines, and individual context \citep{timmermans2002time,
cagney2020urban}. It is also consistent with empirical COVID-19 mobility studies
showing that restrictions changed home activity, destination choice, public
transport use, and discretionary visits in different ways
\citep{Beckinsights2020, yabe2020non, lucchini2021living}. The contribution
here is to turn that insight into a simulation object. Instead of using a single
mobility index, UrbanShare-MoE-PA reconstructs how each agent-day is divided
among home and non-home POI categories and among travel modes.

This distinction is important for both epidemic and activity interpretation.
Two behavioral trajectories can have similar total travel shares but different
exposure and activity implications if one reallocates time toward retail and
public transport while another reallocates time toward home and private car. The
factual reconstruction results show that the model can recover the broad
collapse-and-rebound pattern of the original policy sequence, while the
phase-level and daily trajectory figures show that the recovery differs across
POI categories and mode categories. This supports the view that policy response
is not a homogeneous mobility shock. It is a redistribution of time across urban
functions.

The comparison among UrbanShare, UrbanShare-MoE, and UrbanShare-MoE-PA also
shows why model structure matters. The structured UrbanShare baseline captures
the broad direction of change, but it tends to smooth over individual
heterogeneity and can distort exposure intensity in downstream SEIR simulation.
MoE routing improves this by separating recurring behavioral regimes across
agent-days and phases, while phase-aware preference alignment improves temporal
coherence near policy transitions. Together, these components support an
agent-level behavior generator with heterogeneous response modes and
phase-consistent rollout for policy scenario simulation.

\subsection{Relation to Transport-Epidemic Modeling}
The second implication concerns the link between transport behavior and epidemic
dynamics. Existing SEIR, metapopulation, network, and transport-epidemic models
have established that movement and co-presence shape epidemic propagation
\citep{Keelingnetworks2005, Salatha2010, Wanginferring2018,
Kniplepidemic2013}. In particular, \citet{liu2022modelling} extended the SEIR
framework for Singapore by considering infection within areas and during MRT
commutes. Our results are consistent with that line of work in one central
respect: exposure should be modeled through urban activity and travel contexts,
not through a fully homogeneous mixing assumption.

The difference is where the behavioral input comes from. Much of the earlier
transport-epidemic literature uses observed flows, commuting structures, or
predefined policy parameters as model inputs. That is appropriate when the goal
is to reconstruct an observed epidemic or test a fixed intervention rule. The
present study instead focuses on an upstream problem: when the intervention
calendar changes, the mobility trajectory itself must be generated. This is why
the paper treats NPI scenario analysis as policy scenario simulation. The SEIR
component is used to compare behaviorally generated trajectories on a common
calibrated 911-agent scale.

The comparison also shows how differences at the behavior layer propagate into
epidemic outputs. Over-contraction of non-home activity produces a strongly
attenuated infectious curve, whereas limited phase differentiation makes
alternative policy calendars appear more similar. The downstream SEIR results
therefore reveal the consequences of behavioral-model differences as well as
the resulting policy outcomes.

\subsection{Policy Timing, Duration, and Activity Trade-Offs}
The scenario comparison reinforces a finding that has appeared repeatedly in
COVID-19 modeling studies: timing matters. Studies of social mixing, lockdown,
contact tracing, quarantine, and government interventions generally show that
early action can reduce epidemic burden, while delay can make later control more
difficult \citep{Premthe2020, Davieseffects2020, Kretzschmarimpact2020,
Sharmaunderstanding2021}. The present results are consistent with that logic.
In the calibrated first-wave setting, the early-lockdown scenario reduces the
main infectious peak, whereas the late-lockdown scenario allows more exposure to
accumulate before restrictions become active.

The duration results add a more policy-specific nuance. Extending lockdown does
not automatically dominate shorter alternatives when the extension occurs after
the main peak has formed. In our simulations, the long-lockdown calendar mainly
affects the post-peak tail, while the early-lockdown calendar affects the
first-wave peak itself. In this first-wave setting, the marginal value of
additional restriction depends on epidemic timing and behavioral response. For
policy makers, the relevant question is therefore not only how strict or how
long a measure is, but whether the measure is aligned with the phase of epidemic
growth.

The activity-index results further show that epidemic control and retained
activity cannot be evaluated on separate behavioral scales. Restrictions change
both the amount of activity and its sectoral composition. This is why the short
lockdown retains the highest weighted activity but permits a modest increase in
epidemic burden. Conversely, early lockdown yields the strongest simulated
epidemic improvement but with a larger activity loss. This trade-off is
consistent with broader evidence that COVID-19 restrictions generated uneven
economic costs across places and sectors \citep{bonaccorsi2020economic,
Beckinsights2020}. The policy value of the proposed framework is that it makes
this trade-off behaviorally explicit: the activity index is computed from the same
POI-category and mode-category trajectories that drive exposure.

\subsection{Limitations and Future Studies}
Several limitations define the scope of these findings. First, the epidemic
results are calibrated on a 911-agent behavioral sample and matched to normalized
trend features rather than raw national case counts. They should therefore be
interpreted as sample-scale, trend-calibrated scenario comparisons, not as
forecasts of Singapore-wide infections. Second, alternative-calendar mobility and
epidemic trajectories cannot be observed directly. The outputs are evaluated
through behavioral plausibility, temporal coherence, and downstream consistency,
but they cannot be validated against ground-truth alternative-calendar outcomes or
against unobserved realized outcomes.

Third, the sector-weighted activity index is intentionally coarse. It uses sector-level
GVA weights and population-mean POI and travel-mode hours. It captures retained
activity at the level of broad urban functions, but it does not measure firm
revenue, employment, remote work productivity, household welfare, or
distributional impacts across income and occupational groups. Fourth, the SEIR
module uses a simplified exposure structure. It is sufficient for comparing
scenario trajectories on a common scale, but it does not replace a full
epidemiological observation model with testing, reporting delays, vaccination,
variant dynamics, or household transmission.

Future research could extend this framework in three directions. The first is
scale: larger mobility panels would allow more reliable estimation of rare
activity categories and subgroup heterogeneity. The second is uncertainty:
scenario rollouts should eventually include confidence intervals over behavioral
generation, epidemic parameters, and sector weights. The third is policy
detail: future models could distinguish telework feasibility, essential-worker
mobility, transit crowding management, and destination-specific reopening rules.
Such extensions would move the framework from relative scenario comparison
toward a richer decision-support tool for pandemic-resilient urban and transport
policy.

\section{Conclusion}
\label{sec:conclusion}
This paper developed \textit{UrbanShare-MoE-PA}, a data-driven agent-level
framework for policy scenario simulation under pandemic intervention calendars.
The framework addresses a gap between transport-behavior modeling and
epidemic-policy analysis. Rather than treating mobility as an aggregate input to
an epidemic model, it first generates daily time-allocation trajectories at the
agent level. Each day is represented through travel share, POI-category
allocation, and travel-mode allocation. These behavioral trajectories are then
linked to a calibrated SEIR simulator and a sector-weighted urban activity
index.

The Singapore first-wave case study shows that this behavioral layer is
important. UrbanShare-MoE substantially improves held-out POI reconstruction relative
to the structured UrbanShare baseline, while UrbanShare-MoE-PA achieves the
lowest travel-mode errors and the clearest alternative-calendar trajectories. The MoE components help
represent heterogeneous behavioral regimes, while phase-aware preference
alignment improves responses around policy transition points. These results
support the central argument of the paper: before comparing epidemic or activity
outcomes under alternative NPI calendars, the model must first generate
coherent behavioral trajectories under those calendars.

The calibrated SEIR and activity analyses provide a policy-oriented
illustration of this argument. On the calibrated 911-agent scale, the
early-lockdown calendar produces the strongest simulated reduction in epidemic
burden, while the late-lockdown calendar performs worst epidemiologically. The
short-lockdown calendar retains the highest weighted activity but permits a
modest increase in epidemic burden. The long-lockdown calendar mainly affects
the post-peak tail and does not dominate the original policy. These findings
suggest that, in this first-wave setting, the timing of restriction is more
important than duration alone. They also show why epidemic control and activity
retention should be evaluated through a common behavioral representation.

Future work can extend the framework with larger mobility panels, uncertainty
quantification for behavioral and epidemic simulation, and richer measures of
employment, remote work, firm-level activity, and distributional welfare. These
extensions would broaden the framework toward more detailed decision support for
pandemic-resilient urban and transport policy.

\appendix

\section{Appendix}
\label{app:notation}

\setcounter{figure}{0}
\renewcommand{\thefigure}{A\arabic{figure}}

\begin{figure*}[p]
\centering
\includegraphics[width=0.78\linewidth]{figs/population_daily_factual_legend.png}

\medskip

\includegraphics[width=0.92\linewidth]{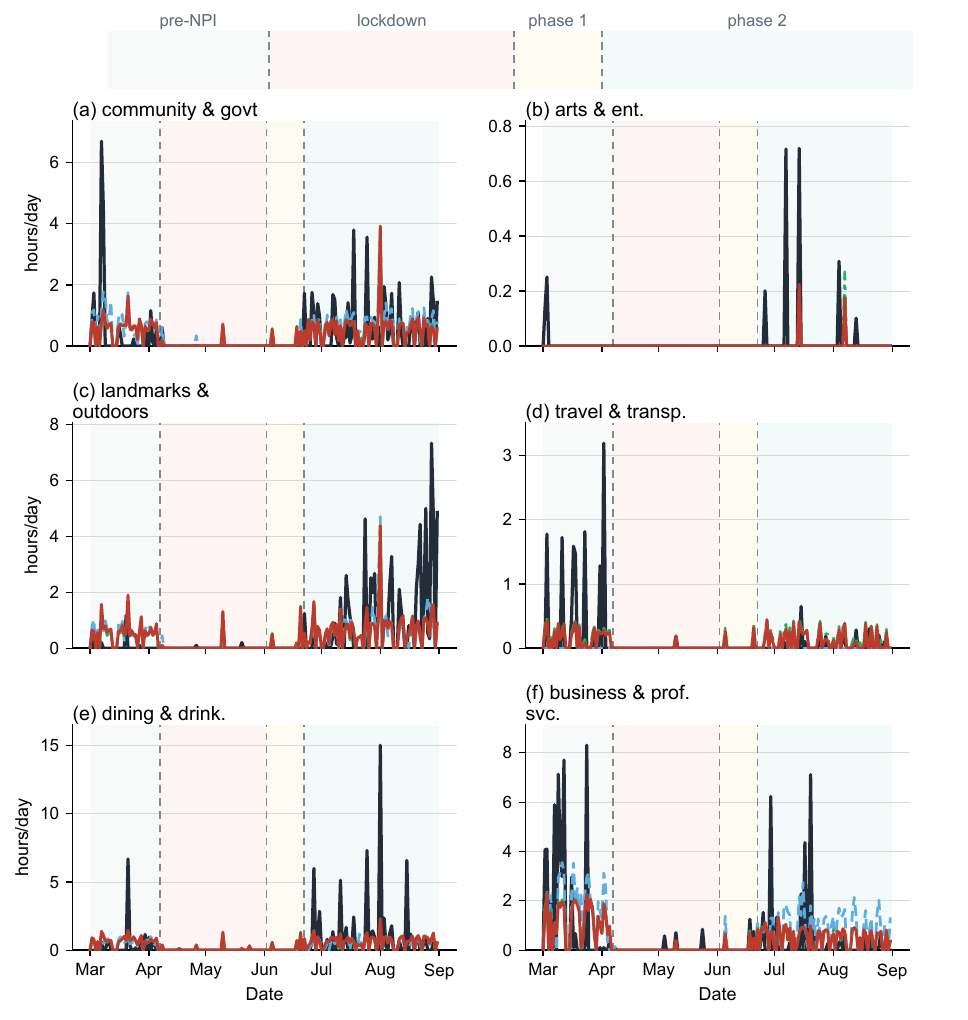}
\caption{Representative-agent POI trajectories (Part~1).}
\label{fig:rep_agent_factual_poi_profiles_a}
\end{figure*}

\begin{figure*}[p]
\centering
\includegraphics[width=0.78\linewidth]{figs/population_daily_factual_legend.png}

\medskip

\includegraphics[width=0.92\linewidth]{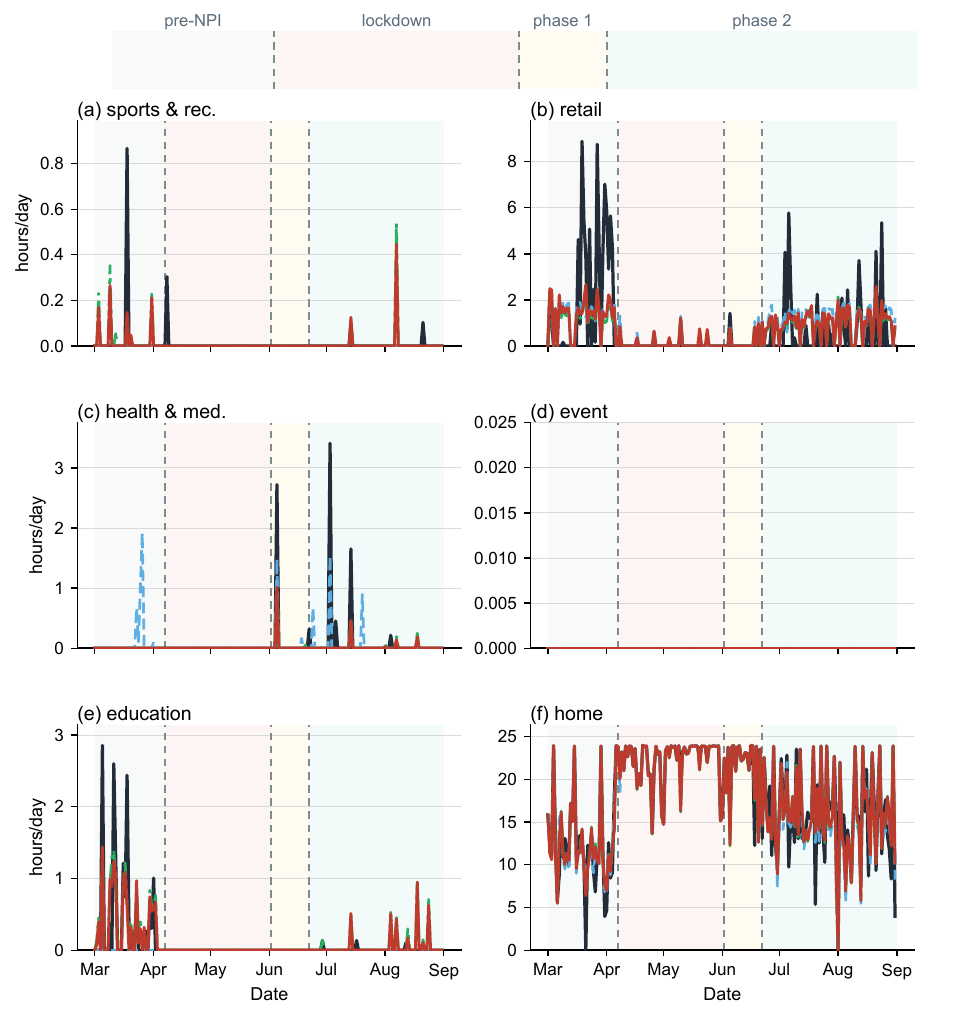}
\caption{Representative-agent POI trajectories (Part~2).}
\label{fig:rep_agent_factual_poi_profiles_b}
\end{figure*}

\begin{figure*}[p]
\centering
\includegraphics[width=0.78\linewidth]{figs/population_daily_factual_legend.png}

\medskip

\includegraphics[width=0.78\linewidth]{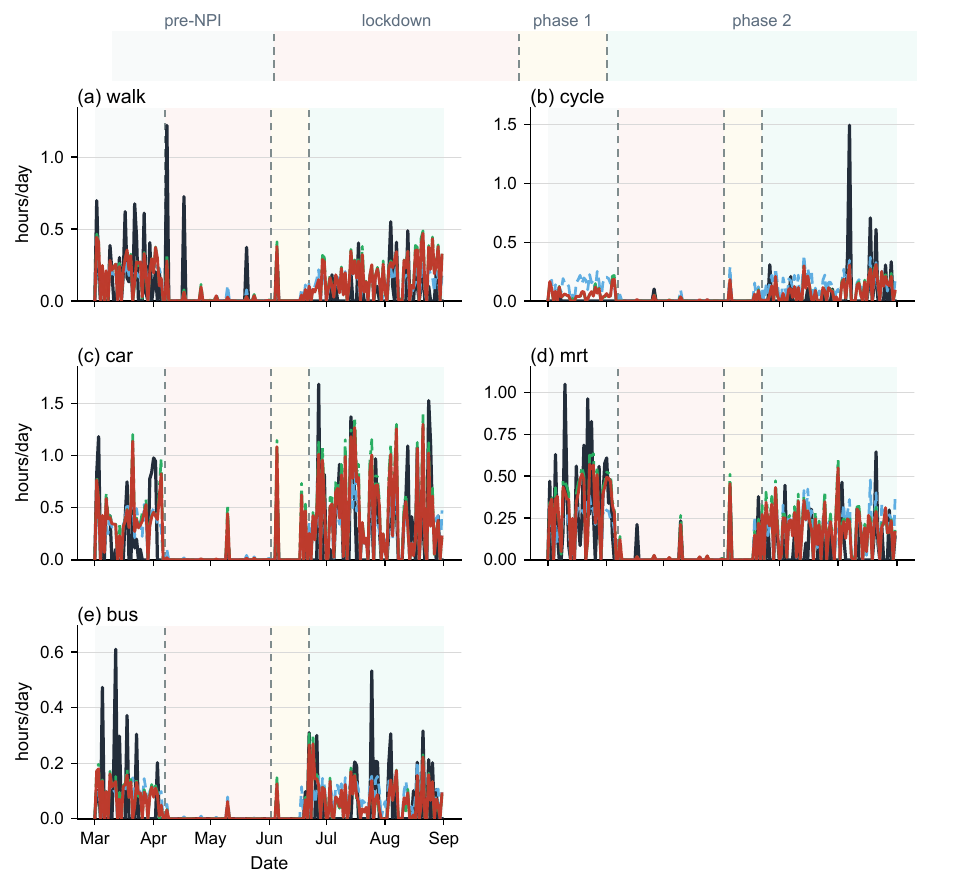}
\caption{Representative-agent travel-mode trajectories.}
\label{fig:rep_agent_factual_mode_profiles}
\end{figure*}

\printcredits

\bibliographystyle{cas-model2-names}

\bibliography{cas-refs}


% Author biography without author photo.
% Author biography. Author biography. Author biography.
% Author biography. Author biography. Author biography.
% Author biography. Author biography. Author biography.
% Author biography. Author biography. Author biography.
% Author biography. Author biography. Author biography.
% Author biography. Author biography. Author biography.
% Author biography. Author biography. Author biography.
% Author biography. Author biography. Author biography.
% Author biography. Author biography. Author biography.

% \bio{figs/pic1}
% Author biography with author photo.
% Author biography. Author biography. Author biography.
% Author biography. Author biography. Author biography.
% Author biography. Author biography. Author biography.
% Author biography. Author biography. Author biography.
% Author biography. Author biography. Author biography.
% Author biography. Author biography. Author biography.
% Author biography. Author biography. Author biography.
% Author biography. Author biography. Author biography.
% Author biography. Author biography. Author biography.

% \bio{figs/pic1}
% Author biography with author photo.
% Author biography. Author biography. Author biography.
% Author biography. Author biography. Author biography.
% Author biography. Author biography. Author biography.
% Author biography. Author biography. Author biography.

\end{document}